\newcommand\epja[3]  {{ Eur.\ Phys.\ J. }{\bf A #1} (#2) #3}
\newcommand\epjc[3]  {{ Eur.\ Phys.\ J. }{\bf C #1} (#2) #3}
\newcommand\jpg[3]   {{ J.\ Phys.\ }{\bf G #1} (#2) #3}

\newcommand\jhep[3]  {{ J. High Energy Phys.\ }{\bf #1} (#2) #3}
\newcommand\npb[3]    {{ Nucl.\ Phys.\ }{\bf B #1} (#2) #3}
\newcommand\npbps[3]  {{ Nucl.\ Phys.\ B\ (Proc.\ Suppl.)\ }{\bf #1} (#2) #3}

\newcommand\plb[3]   {{ Phys.\ Lett.\ }{\bf B #1} (#2) #3}
\newcommand\ppnp[3]  {{ Prog.\ Part.\ Nucl.\ Phys.\ }{\bf #1} (#2) #3}

\newcommand\prd[3]   {{ Phys.\ Rev.\ }{\bf D #1} (#2) #3}
\newcommand\prep[3]  {{ Phys.\ Rep.\ }{\bf #1} (#2) #3}
\newcommand\prl[3]   {{ Phys.\ Rev.\ Lett.\ }{\bf #1} (#2) #3}
\newcommand\ptp[3]   {{ Prog.\ Theor.\ Phys.\ }{\bf #1} (#2) #3}

\newcommand\mpla[3]  {{ Mod.\ Phys.\ Lett.\ }{\bf A #1} (#2) #3}

\newcommand\jetp[3]  {{ Sov.\ Phys.\ JETP\/ }{\bf #1} (#2) #3}

\newcommand\zetf[3]  {{ Zh.\ Eksp.\ Teor.\ Fiz.\ }{\bf #1} (#2) #3}

\newcommand\ijmpa[3] {{ Int.\ J.\ Mod.\ Phys.\ }{\bf A #1} (#2) #3}

\newcommand{\hepph}[1] {{\tt hep-ph/#1}}

\newcommand{\hepex}[1] {{\tt hep-ex/#1}}

\documentclass{elsart}
\usepackage{amssymb,amsmath,graphicx}

\begin{document}
\def \meff{\mbox{$\left|  < \! m  \! > \right| \ $}}
\def\rsol{r_{\odot}}
\def\dmsol{\Delta m^2_\odot}
\def\dmatm{\Delta m^2_a}
\def\tgatm{\tan^2\theta_a}
\def\tgsol{\tan^2\theta_\odot}
\def\nl{\nonumber \\}

\def\openone{\leavevmode\hbox{\small1\kern-3.3pt\normalsize1}}

\begin{frontmatter}
\begin{flushright}
FISIST/03\,-2002/CFIF \\
\end{flushright}

\title{Leptogenesis, $CP$ violation and neutrino data: What can we
learn?}
\author{G. C. Branco},
\author{R. Gonz\'{a}lez Felipe},
\author{F. R. Joaquim},
\author{M. N. Rebelo}
\address{{\emph Centro de F\'{\i}sica das Interac\c{c}\~{o}es Fundamentais
(CFIF), Departamento de F\'{\i}sica, Instituto Superior T\'{e}cnico, Av.
Rovisco Pais, 1049-001 Lisboa, Portugal}}
\thanks{E-mails: gbranco@cfif.ist.utl.pt (G.C. Branco),
gonzalez@gtae3.ist.utl.pt (R. Gonz\'{a}lez Felipe),
filipe@gtae3.ist.utl.pt (F.R. Joaquim), rebelo@alfa.ist.utl.pt (M.N.
Rebelo)}

\begin{abstract}
A detailed analytic and numerical study of baryogenesis through leptogenesis is
performed in the framework of the standard model of electroweak interactions
extended by the addition of three right-handed neutrinos, leading to the seesaw
mechanism. We analyze the connection between GUT-motivated relations for the
quark and lepton mass matrices and the possibility of obtaining a viable
leptogenesis scenario. In particular, we analyze whether the constraints
imposed by SO(10) GUTs can be compatible with all the available solar,
atmospheric and reactor neutrino data and, simultaneously, be capable of
producing the required baryon asymmetry via the leptogenesis mechanism. It is
found that the Just-So$^2$ and SMA solar solutions lead to a viable
leptogenesis even for the simplest SO(10) GUT, while the LMA, LOW and VO solar
solutions would require a different hierarchy for the Dirac neutrino masses in
order to generate the observed baryon asymmetry. Some implications on $CP$
violation at low energies and on neutrinoless double beta decay are also
considered.
\end{abstract}
%\begin{keyword}
%Leptogenesis; neutrino masses and mixing; $CP$ violation.
%\end{keyword}
\end{frontmatter}

\section{Introduction}
\label{sec1}

Baryogenesis via leptogenesis is one of the most appealing mechanisms for
generating the observed baryon asymmetry of the Universe (BAU). The interest in
leptogenesis has been reinforced after the strong evidence for neutrino
oscillations reported by the Super-Kamiokande experiment \cite{SK} and recently
confirmed by the results of the Sudbury Neutrino Observatory (SNO) \cite{SNO},
both pointing towards nonzero neutrino masses. From a theoretical point of
view, the smallness of neutrino masses can be naturally explained through the
seesaw mechanism \cite{seesaw} which automatically follows, once right-handed
neutrinos are added to the standard model (SM). Since the right-handed neutrino
fields are singlets under the SU(3)$_c$ $\times$ SU(2) $\times$ U(1) gauge
symmetry of the SM, right-handed neutrino Majorana masses are not protected by
the gauge symmetry so they can be much larger than the electroweak scale. These
right-handed neutrino states appear naturally in Grand Unified Theories (GUT)
such as SO(10) \cite{so10}, where they are included in the same irreducible
representation together with quarks and leptons.

At low energies, all the information about neutrino masses and mixing is
contained in the effective neutrino mass matrix which can be expressed by the
seesaw formula:
\begin{align} \label{intro1}
M_\nu \simeq-M_D M_R^{-1} M_D^T\ ,
\end{align}
where $M_D$ and $M_R$ are the Dirac and right-handed neutrino mass matrices,
respectively. Unfortunately,  $M_D$ and $M_R$ are not fully determined by the
low-energy experimental data. In this respect, GUTs can play an essential
r\^ole since in their framework $M_D$ is closely related to the quark or
charged lepton mass matrices. Another alternative to this approach is to impose
some extra symmetries like horizontal or discrete symmetries \cite{extra}.

In this paper we investigate the constraints on $M_D$ resulting from the
requirement of a viable leptogenesis \cite{buchmuller,textures,hirsch}. More
specifically, we investigate whether leptogenesis can discriminate among the
various solar neutrino solutions in the framework of some GUT-inspired patterns
for $M_D$. We shall assume a special form for $M_D$ in the weak basis where
$M_R$ and the charged leptonic mass matrix $M_\ell$ are diagonal. There is of
course no loss of generality in choosing these two matrices real, positive and
diagonal. Since our discussion does not rely on specific textures for these
matrices it will cover a wide range of mass matrices. In this sense our
analysis is quite general. Yet, as we shall see, the hierarchy of the masses
resulting from $M_D$ plays a crucial r{\^o}le in the viability of the leptogenesis
scenario. In fact, mass hierarchies such as the ones imposed by minimal SO(10)
strongly constrain the allowed solar solutions. Nevertheless, it turns out that
allowing for a more general choice of the mass spectrum of $M_D$, it is
possible to reconcile the different solar solutions with the required
cosmological BAU.

The connection between neutrino masses and mixing and the observed baryon
asymmetry of the Universe becomes clear if one takes into account that the
leptogenesis mechanism starts with the $CP$ asymmetry generated through the
out-of-equilibrium $L$-violating decays of the heavy Majorana neutrinos
\cite{fukugita,luty,buchmuller}, leading to a lepton asymmetry which is
subsequently transformed into a baryon asymmetry by the $(B+L)$-violating
sphaleron processes \cite{kuzmin}. As it will be discussed later, this baryon
asymmetry depends mainly on the heavy Majorana neutrino mass spectrum and the
Dirac neutrino Yukawa coupling matrix, in the basis where the mass matrix $M_R$
is diagonal. These matrices will be constrained by GUT-like relations and by
the available information on neutrino mixing and mass squared differences
measured in solar, reactor and atmospheric neutrino experiments. On the other
hand, requiring the produced baryon-to-entropy ratio $Y_B \equiv n_B/s$  to be
in the presently allowed range \cite{PDG},
\begin{align} \label{intro2}
1.7 \times 10^{-11} \lesssim Y_B \lesssim 8.1 \times 10^{-11}\ ,
\end{align}
can in principle constrain some mixing and mass parameters, or even
discriminate the solar neutrino solutions which are compatible with the
leptogenesis scenario. We shall see that the only parameters relevant to this
analysis are the absolute value of the (1,3)-element of the leptonic mixing
matrix, the Dirac-type phase and the two Majorana phases appearing in this
matrix, as well as the mass of the lightest neutrino.

At this stage one may wonder whether there is any link between leptogenesis and
leptonic $CP$ violation at low energies since they both arise from the phases
appearing in the leptonic mixing matrix. Several authors have already addressed
this question under different assumptions \cite{link,branco}. Experimentally,
leptonic $CP$-violating effects can be probed in neutrino oscillation
experiments and they are only sensitive to the Dirac-type phase appearing in
the leptonic mixing matrix. On the other hand, the size of double beta decays
is affected by the presence of Majorana-type $CP$-violating phases.

The paper is organized as follows. In Section~\ref{sec2} we present the general
framework and the strategy to follow. Special emphasis is given to the
construction of the Dirac, Majorana and effective neutrino mass matrices as
well as to the identification of the $CP$-violating phases relevant to
leptogenesis. In Section~\ref{sec3} we present the simplest leptogenesis
scenario, namely, the out-of-equilibrium decay of a single heavy Majorana
neutrino. We briefly recall how to estimate the baryon asymmetry taking into
account the washout effects due to inverse decays and lepton number violating
scattering processes. This section is divided into two parts. First we develop
a simple analytic approach which will allow us to obtain upper bounds on the
baryon asymmetry for the different solar neutrino solutions and assuming
different patterns for the light neutrino spectrum. We then perform a detailed
numerical analysis in order to identify the regions of the parameter space
where the leptogenesis mechanism is efficient enough to produce the required
cosmological baryon asymmetry, taking into account the presently available low
energy neutrino data. Section \ref{sec4} is devoted to discuss some
implications on leptonic $CP$ violation at low energies and neutrinoless double
beta decay. Finally, our conclusions are presented in Section~\ref{sec5}.

\section{General framework}
\label{sec2}

We shall work in the framework of a minimal extension of the SM with only one
right-handed Majorana neutrino per generation. After spontaneous symmetry
breakdown, the leptonic mass terms are of the form
\begin{align} \label{gf1}
 -\mathcal{L}_{mass}=\bar{\ell}_L^{\;0}\,M_\ell\,\ell_R^{\;0}
+\bar{\nu}_L^0\,M_D\, \nu_R^0+ \frac{1}{2}\,\nu_R^{0\;T} \,C\,M_R\,
\nu_R^0+\text{h.c.}\,,
\end{align}
where ${\ell}_{L,R}^{\;0}$ and $\nu_{L,R}^0$ are the charged lepton and
neutrino weak eigenstates, respectively. The charged lepton mass matrix
$M_\ell$ and the Dirac neutrino mass matrix $M_D$ are in general $3 \times 3$
complex matrices whilst the heavy Majorana neutrino mass matrix $M_R$ is
constrained to be symmetric. The right-handed Majorana mass term is
SU(2)$\times$U(1) invariant, as a result it is the only fermionic mass term
present in the Lagrangian before the symmetry breaking. The Lagrangian in
Eq.~(\ref{gf1}) can be rewritten as
\begin{align} \label{gf2}
-\mathcal{L}_{mass}=\frac{1}{2}\,n_L^T
C \mathcal{M}^{*}n_L+\bar{\ell}_L^{\;0}\,M_\ell\,\ell_R^{\;0}+\text{h.c.}\,,
\end{align}
with $n_L=\left(\nu_L^0,(\nu_R^0)^C \right)$ and
\begin{align} \label{gf3}
\mathcal{M}=\left(
\begin{array}{cc}
 0     &\quad M_D\\
 M_D^T &\quad M_R
\end{array}\right)\,.
\end{align}
The zero entry in $\mathcal{M}^{*}$ is due to the absence, at tree level, of a
Majorana mass term for the left-handed neutrinos $\nu_L$. The diagonalization
of $\mathcal{M}^{*}$ is performed through the transformation
\begin{align} \label{gf4}
V^T\,\mathcal{M}^{*}\,V=\mathcal{D}=\text{diag}(\,m_1,m_2,m_3,M_1,M_2,M_3)\,,
\end{align}
where $m_i$ and $M_i$ denote the physical masses of the light and heavy
Majorana neutrinos, respectively. It is useful to write $V$ and $\mathcal{D}$
in a block form as
\begin{align} \label{gf5}
V=\left(
\begin{array}{cc}
 K     &\quad R\\
 S &\quad T
\end{array}\right)\quad,\quad
\mathcal{D}=\left(
\begin{array}{cc}
 d_\nu     &\quad 0\\
 0 &\quad d_R
\end{array}\right)\,,
\end{align}
with $d_\nu=\text{diag}(m_1,m_2,m_3), d_R=\text{diag}(M_1,M_2,M_3)$. From
Eq.~(\ref{gf4}) one derives the approximate relations
\begin{align} \label{gf6}
-K^{\dagger}M_D M_R^{-1} M_D^T K^{*}\equiv
K^{\dagger}M_\nu K^{*}=d_\nu\,,
\end{align}
\begin{align} \label{gf7}
S^{\dagger}=-K^{\dagger}M_D M_R^{-1} \,,
\end{align}
together with the exact relation
\begin{align} \label{gf8}
R=M_D\,T^{*}\,d_R^{-1}\,.
\end{align}
The relation in Eq.~(\ref{gf6}) is the usual seesaw formula, which leads in a
natural way to small left-handed Majorana neutrino masses of order $M_D^2
M_R^{-1}$. The matrix $K$ is almost unitary, which is in agreement with the
matrices $S$ and $R$ being of order $M_D\,M_R^{-1}$ as can be inferred from
Eqs.(\ref{gf7}) and (\ref{gf8}).

The neutrino weak eigenstates $\nu_{iL}^0$ are related to the mass eigenstates
by
\begin{align} \label{gf9}
\nu_{iL}^0=V_{i\alpha} \nu_{\alpha L} = (K,R) \left(
\begin{array}{c}
\nu_{iL} \\ N_{iL}
\end{array} \right)\ ,\quad i=1,2,3\ ,\quad \alpha = 1,2,\cdots,6\ ,
\end{align}
and thus the leptonic charged current interactions are given by
\begin{align} \label{gf9a}
-\frac{g}{\sqrt{2}} \left( \bar{\ell}_L\, \gamma_\mu\, K\, \nu_L +
\bar{\ell}_L\, \gamma_\mu\, R\, N_L \right) W^\mu +\text{h.c.}\,,
\end{align}
so that $K$ and $R$ give the charged current couplings of the charged leptons
to the light neutrinos $\nu_i$ and the heavy ones $N_i$, respectively. Without
imposing any restriction on the mass matrices, Eq.~(\ref{gf1}) can be rewritten
in the weak basis (WB) where $M_\ell$ and $M_R$ are real and diagonal in the
form
\begin{align} \label{gf10}
 -\mathcal{L}_{mass}=\bar{\ell}_L^{\;0}\,d_\ell\,\ell_R^{\;0}
+\bar{\nu}_L^0\,M_D\, \nu_R^0+ \frac{1}{2}\,\nu_R^{0\;T} \,C\,d_R\,
\nu_R^0+\text{h.c.}\,,
\end{align}
where $d_\ell = \text{diag}(m_e,m_\mu,m_\tau)$. We have kept for simplicity the
same notation of Eq.~(\ref{gf1}) for the rotated fields and mass matrices in
the Lagrangian (\ref{gf10}). Obviously, in this WB all $CP$-violating phases
appear in $M_D$. Since for $n$ generations there are $n (n-1)$ independent
phases \cite{branco,endoh}, the matrix $M_D$ has six $CP$-violating phases. In
addition, the Lagrangian $\mathcal{L}_{mass}$ has fifteen real physical
parameters: nine contained in $M_D$ plus the six masses in $d_\ell$ and $d_R$.

It is easy to see how these phases appear in $M_D$. Using the polar
decomposition, $M_D$ can be written as the product of a unitary matrix $U$
times a Hermitian matrix $H$. Factoring out all possible phases we can write
\begin{align} \label{gf11}
M_D = U H = P_\xi U' P_1 H' P_2\ ,
\end{align}
with $P_\xi= \text{diag}(e^{i \xi_1}, e^{i \xi_2}, e^{i \xi_3}),
P_j=\text{diag}(1, e^{i \varphi_1^j}, e^{i \varphi_2^j})$ and the matrices $U'$
and $H'$ containing only one phase each. Since $P_\xi$ can always be rotated
away by a simultaneous phase transformation of the left-handed charged lepton
fields and left-handed neutrino fields, only six phases are physically
meaningful.

In the physical basis all $CP$-violating phases are shifted to the leptonic
mixing matrix and thus will appear in the matrices $K$ and $R$. Since to a very
good approximation $K$ is unitary and three of its phases can be rotated away,
we end up with three physical phases in $K$, two of which are of Majorana
character. These phases are in general complicated functions of the six phases
of $M_D$ and can be obtained from Eq.~(\ref{gf6}) by replacing $M_R$ by $d_R$.
The matrix $R$, on the other hand, verifies Eq.~(\ref{gf8}). In the WB where
$M_R$ is real and diagonal, the matrix $T$ in (\ref{gf5}) is close to the
identity. As a result, from Eqs.~(\ref{gf8}) and (\ref{gf11}) we obtain
\begin{align} \label{gf12}
R = U'P_1 H' d_R^{-1} P_2\ .
\end{align}
Once again two of the phases appearing in $R$ are of Majorana character - the
ones contained in $P_2$. As we shall see in the next section, leptogenesis is
only sensitive to the phases appearing in $M_D^\dagger M_D$, or equivalently
$R^\dagger R$. Therefore, from Eq.~(\ref{gf12}) we can conclude that the only
$CP$-violating phases relevant for leptogenesis are the phase contained in $H'$
and the two Majorana phases of $P_2$.

In the exact decoupling limit, $R$ can be neglected and only $K$ is relevant.
In this limit, the leptonic mixing matrix $K$ is usually referred as the
Pontecorvo-Maki-Nakagawa-Sakata (PMNS) matrix \cite{pontecorvo,maki}, which
from now on we shall denote as $U_\nu$. However, since we are interested in
studying the connection between $CP$ violation at low energies and leptogenesis
we should keep both $K$ and $R$ or, alternatively, the Dirac neutrino mass
matrix $M_D$ in the WB where $M_\ell$ and $M_R$ are diagonal.

Experimentally, only the charged lepton mass spectrum is very well known. In
contrast, the experimental data on light neutrino masses and leptonic mixing
define different allowed regions in parameter space, still leaving a lot of
freedom in the choice of patterns for the matrices $U_\nu$ and $M_\nu$. The
heavy neutrino masses and their mixing with the charged leptons are even less
constrained.

From the theoretical point of view, grand unified theories such as SO(10)
\cite{so10} are a suitable framework not only to analyze fermion masses but
also to implement the seesaw mechanism. One of the attractive features of the
SO(10) model is that its gauge group is left-right symmetric and, consequently,
there exists a complete quark-lepton symmetry in the spectrum. In particular,
the fact that all left-handed (right-handed) fermions of each family fit into
the single irreducible spinor representation 16~($\overline{16}$) of SO(10) and
that the right-handed neutrino is precisely contained in this representation is
remarkable.  Several constraints on fermion masses are usually implied in these
models \cite{so10}. For instance, if there is only one 10 Higgs multiplet
responsible for the masses, then we have the relation $M_u=M_d=M_\ell=M_D $
(the indices $u$ and $d$ stand for the up and down quarks, respectively).
Similarly, the existence of two 10 Higgs multiplets implies $M_\ell=M_d$ and
$M_D=M_u $. On the other hand, if the fermion masses are generated by a VEV of
the 126 of SO(10), then the SU(4) symmetry yields the relations $3\,
M_d=-M_\ell$ and $3\, M_u=-M_D $. Of course, equalities such as the ones
arising purely from the 10-dimensional representation cannot be exact in
realistic models, since they imply unphysical relations among the quark and
charged lepton masses. Additional assumptions are therefore necessary in order
to predict the correct fermion spectrum \cite{georgi,ramond,babu}.

Without loss of generality, one can choose a WB where $M_\ell$ and $M_R$ are
diagonal and real while $M_D$ is an arbitrary complex matrix which we write as:
\begin{align} \label{gf13}
M_D=V_L^{\dag}\,d_D\,U_R\ ,
\end{align}
where  $d_D =\text{diag} (m_{D1}, m_{D2}, m_{D3})$ and $V_L$, $U_R$ are unitary
matrices. In a minimal SO(10) scenario it is  expected a small misalignment
between the Hermitian matrices $M_\ell M_\ell^\dagger$ and $M_D M_D^\dagger$,
similar to that in the quark sector. The matrix $V_L$ is analogous to the
Cabibbo-Kobayashi-Maskawa (CKM) matrix of the quark sector and under the above
assumption of small misalignment, it should be close to the identity matrix.
The Dirac neutrino mass spectrum will be constrained by SO(10) relations such
as the ones discussed above. For definiteness, we will assume that $d_D$ is
given by the up-quark spectrum, i.e
\begin{align} \label{gf14}
d_D = \text{diag}(m_u, m_c, m_t)\ .
\end{align}
A brief discussion of the most general case with arbitrary masses is presented
at the end of Section \ref{sec3b}.

In our analysis we choose $d_\nu$ and $U_\nu$ in agreement with the present
experimental data. The corresponding effective neutrino matrix $M_\nu$ can be
then computed from Eq.~(\ref{gf6}), i.e.
\begin{align} \label{gf15}
M_\nu = U_\nu\, d_\nu\, U_\nu^T\ .
\end{align}
We shall consider different patterns for the PMNS matrix $U_\nu$ corresponding
to the different solar solutions (LMA, SMA, LOW, VO, Just-So$^2$). For each
solution we impose the corresponding mass squared differences $\dmsol \equiv
\Delta m_{12}^2 = |m^2_2 - m^2_1|$ (solar neutrinos) and $\dmatm \equiv \Delta
m_{23}^2 = |m^2_3 - m^2_2|$ (atmospheric neutrinos) to be in the presently
allowed experimental range. We also let the mass of the lightest neutrino to
vary in the appropriate range: its lower bound is typically fixed by requiring
the heaviest Majorana mass to be below the Planck scale, while the upper bound
is chosen so that neutrinoless double beta decays experiments are not violated.

An important feature of this approach is that once $M_\nu$ is chosen, and for a
given\footnote{Throughout our paper we either take $V_L$ equal to the identity
or close to it.} $V_L\ $, the masses of the heavy Majorana neutrinos are fixed
together with the matrix $U_R$, which appears in Eq.~(\ref{gf13}). Indeed, the
relations
\begin{align} \label{gf16}
M_{\nu} &= -M_D\,d_R^{-1}\,{M_D}^{T} = -V_L^\dagger\,
d_D\, M\, d_D\, V_L^*\ ,\nl
M &\equiv -d_D^{-1}\,V_L\,M_{\nu}\,{V_L}^{T}\,d_D^{-1} =
U_R\, d_R^{-1}\, {U_R}^T \ ,
\end{align}
imply that the matrix $M$ is fully determined. Obviously, we have $
M\,M^\dagger = U_R\, d_R^{-2}\,U_R^\dagger\ $, corresponding to an eigenvalue
equation. This equation determines the heavy Majorana spectrum $d_R$ and also
the unitary matrix $U_R$ up to a phase ambiguity, which can be then fixed by
going back to Eqs.~(\ref{gf16}).

Notice that Eqs.~(\ref{gf16}) clearly illustrate a remarkable feature of these
kind of models \cite{altarelli}, which is the fact the large mixing required
for the atmospheric solution as well as some of the solar solutions can be
obtained with small mixing in $U_R$, so that nearly maximal neutrino mixing is
produced through the seesaw mechanism. In other words, starting from nearly
diagonal mass matrices and as a result of the interplay between the small
off-diagonal entries of $U_R$ and a strong hierarchy in $d_D$ and $d_R$, it is
possible to obtain a large neutrino mixing via the seesaw mechanism.

\section{Heavy Majorana neutrino decays, leptogenesis and neutrino
oscillations}
\label{sec3}

The crucial ingredient in leptogenesis scenarios is the $CP$ asymmetry
generated through the interference between tree-level and one-loop heavy
Majorana neutrino decay diagrams. In the simplest extension of the SM, such
diagrams correspond to the decay of the Majorana neutrino into a lepton and a
Higgs boson. Considering the decay of one heavy Majorana neutrino $N_i$, the
$CP$ asymmetry in the SM is then given by
\begin{align} \label{lepto0}
\epsilon_{N_i}=\frac{\Gamma\,(N_i \rightarrow l\,H)-\Gamma \,(N_i
\rightarrow \bar{l}\,H^{\ast})}{\Gamma\,(N_i \rightarrow
l\,H)+\Gamma\,(N_i \rightarrow \bar{l}\,H^{\ast})}\ .
\end{align}

If the heavy Majorana neutrino masses are such that $M_1 < M_2 < M_3$ only the
decay of the lightest Majorana neutrino $N_1$ is relevant for the lepton
asymmetry\footnote{This is a reasonable assumption if the interactions of the
lightest Majorana neutrino $N_1$ are in thermal equilibrium at the time of the
$N_{2,3}$ decays, so that the asymmetries produced by the heaviest neutrino
decays are erased before the lightest one decays, or if $N_{2,3}$ are too heavy
to be produced after inflation \cite{giudice}.} and one obtains
\cite{covi,buchmuller}
\begin{align} \label{lepto1}
\epsilon_{N_1}=\frac{1}{8\,\pi\,v^2}
\frac{1}{({M_D}^{\dag}M_D)_{11}}\,\sum_{i=2,3} {\rm Im}\left[
 ({M_D}^{\dag}M_D)_{1i}^{2}\right]
 \left[f\!\left(\frac{M_i^2}{M_1^2}\right)+g\!\left(\frac{M_i^2}{M_1^2}\right)
 \right]\ ,
\end{align}
where $M_D$ is the Dirac neutrino mass matrix in the basis where $M_R$ is
diagonal  and $v=\langle H_0 \rangle /\sqrt{2} \simeq 174\,$GeV. The functions
$f(x)$ and $g(x)$ denote the one-loop vertex and self-energy corrections,
respectively, and are given by
\begin{align} \label{lepto2}
 f(x)=\sqrt{x}\left[ 1+
(1+x)\ln\left(\frac{x}{1+x}\right)\right] \quad,\quad
g(x)=\frac{\sqrt{x}}{1-x}\,.
\end{align}
In the limit $M_1 \ll M_2 , M_3 $ the $CP$ asymmetry (\ref{lepto1})
is approximately given by
\begin{align} \label{lepto3}
\epsilon_{N_1}\simeq
-\frac{3}{16\,\pi v^2}\,
\left(I_{12}\,\frac{M_1}{M_2} + I_{13}\,\frac{M_1}{M_3}\right)\,,
\end{align}
where
\begin{align} \label{lepto4}
I_{1i} \equiv \frac{{\rm Im}\left[
 (M_D^\dagger M_D)_{1i}^2 \right]}{ (M_D^\dagger \,M_D)_{11}}\ .
\end{align}
The lepton asymmetry $Y_L$ is related to the $CP$ asymmetry through the
relation
\begin{align} \label{lepto5}
Y_L=\frac{n_L-n_{\bar{L}}}{s}=d\,\frac{\epsilon_{N_1}}{g_{\ast}}\
,
\end{align}
where $g_{\ast}$ is the effective number of relativistic degrees of freedom
contributing to the entropy and $d$ is the so-called dilution factor which
accounts for the washout processes (inverse decay and lepton number violating
scattering). In the SM case, $g_{\ast}=106.75$.

The produced lepton asymmetry $Y_L$ is converted into a net baryon asymmetry
$Y_B$ through the $(B+L)$-violating sphaleron processes. One  finds the
relation \cite{harvey,buchmuller}
\begin{align} \label{lepto6}
Y_B=\xi\,Y_{B-L}=\frac{\xi}{\xi-1}\,Y_L\;\;,\;\;\xi=
\frac{8\,N_f+4\,N_H}{22\,N_f+13\,N_H}\,,
\end{align}
where $N_f$ and $N_H$ are the number of fermion families and complex Higgs
doublets, respectively. Taking into account that $N_f=3$ and $N_H=1$ for the
SM, we get $\xi \simeq 1/3$ and
\begin{align}  \label{lepto7}
Y_B \simeq -\frac{1}{2}\,Y_L\,.
\end{align}

The determination of the dilution factor involves the integration of the full
set of Boltzmann equations. A simple approximated solution which has been
frequently used is given by \cite{kolb}
\begin{align}
d =\left\{
\begin{array}{l}  \label{lepto8}
\sqrt{\,0.1\,\kappa}\,\exp\left(-\frac{4}{3}\sqrt[4]{\,0.1\,\kappa}\right)
\quad , \quad  \kappa \gtrsim 10^{6}  \\
 0.24 (\kappa\,\ln \kappa)^{-3/5}\quad , \quad
10\lesssim \kappa \lesssim 10^6 \\
1/(2 \kappa) \quad , \quad 1\lesssim \kappa \lesssim 10 \\
1  \quad , \quad 0\lesssim \kappa \lesssim 1
\end{array}
\right.
\end{align}
where the parameter $\kappa$, which measures the efficiency in producing the
asymmetry, is defined as the ratio of the thermal average of the $N_1$ decay
rate and the Hubble parameter at the temperature $T = M_1\ $,
\begin{align} \label{lepto9}
\kappa=\frac{M_P}{1.7 \times 8 \pi v^2  \sqrt{g_{\ast}}}
\frac{({M_D}^{\dag}\,M_D)_{11}}{M_1}\ ,
\end{align}
$M_P \simeq 1.22 \times 10^{19}$~GeV is the Planck mass\footnote{Another
variable commonly used in the literature \cite{buchmuller,giudice,hirsch} is
the mass parameter $\tilde{m}_1 = (M_D^\dag\,M_D)_{11}/M_1 \simeq 1.1 \times
10^{-3} \kappa$~eV.}. A slightly modified approximation is used instead by the
authors of Ref.~\cite{nielsen}:
\begin{align}
d =\left\{
\begin{array}{l}  \label{lepto8a}
 0.30 (\kappa\,\ln \kappa)^{-3/5}\quad , \quad
10\lesssim \kappa \lesssim 10^6 \\
1/(2 \sqrt{\kappa^2+9}\,) \quad , \quad 0 \lesssim \kappa \lesssim 10
\end{array}
\right.
\end{align}

However, it has been recently pointed out \cite{hirsch} that in some cases the
above approximations could seriously underestimate the suppression in the
baryon asymmetry due to the washout effects. A more reliable result is obtained
if one uses the empirical fit \cite{hirsch}
\begin{align} \label{lepto10}
&\log_{10} d = \log_{10} (1 - \xi) + {\rm min}\{d_1, d_2, d_3\}\ , \nl
&d_1 = 0.8\, \log_{10} \kappa
-0.7 + 0.05\, \log_{10} M_1^{10}\ , \\
&d_2 = -1.2 - 0.05\, \log_{10} M_1^{10}\ , \nl
&d_3 =  -(3.8 + \log_{10} M_1^{10})
(\log_{10} \kappa - 1) - (5.4 -0.67\,\log_{10} (M_1/{\rm GeV}))^2
-1.5 \ , \nonumber
\end{align}
where $\xi$ is defined in Eq.(\ref{lepto6}), $M_1^{10} \equiv M_1/10^{10}\,
{\rm GeV}$ and the function \emph{min} evaluates to the smallest of the
quantities $d_i$. This fit reproduces considerably better the exact solution
\cite{buchmuller} of the Boltzmann equations for a wider range of $\kappa$ and
$M_1$. We notice however that for $M_1 < 10^8$~GeV and $\kappa \gg 1$ the
approximation (\ref{lepto8a}) tends to give asymptotically a better result than
the empirical fit in Eq.~(\ref{lepto10}) \cite{hirsch:priv}.

We shall therefore use a combined fit in our analysis in order to estimate the
washout effects. If $M_1 \gtrsim 10^8$~GeV or $M_1 < 10^8$~GeV with $\kappa \ll
1$, we calculate $d$ using Eqs.~(\ref{lepto10}). On the other hand, if $M_1 <
10^8$~GeV and $\kappa \gtrsim 1$ we make use of the approximation
(\ref{lepto8a}). This will allow us to obtain a simple and reliable result for
the solution of the Boltzmann equations without resorting to the full numerical
solution of these equations.

\subsection{Analytic approach}
\label{sec3a}

In this section we present a simple analytic approach that will allow us to
obtain upper bounds for the baryon asymmetry in the present framework for the
different solar neutrino solutions and assuming different patterns for the
light neutrino spectrum. We shall divide our analysis in two parts. First we
consider the case of large solar mixing, which includes four solar solutions:
LMA, LOW, VO and Just-So$^2$ solar solutions. Secondly, we discuss the case of
small solar mixing, i.e. the SMA solar solution. We shall consider for each
case two possible patterns for the light neutrino spectrum: hierarchical, i.e.
$m_1 \ll m_2 \ll m_3$ and inverted-hierarchical, $m_1 \sim m_2 \gg m_3\ $, with
the mass squared differences corresponding to the observed hierarchies $\dmatm
\gg \dmsol$. The analysis of another possible pattern, namely the case when the
light neutrino masses are almost degenerate, $m_1 \sim m_2 \sim m_3\ $,  is
more subtle and crucially depends on the inclusion of the $CP$-violating phases
in the leptonic mixing matrix. Therefore we shall not discuss it here. However,
it turns out from our full numerical study of Section \ref{sec3b} that in the
latter case the produced baryon asymmetry is highly suppressed for all the
solar solutions.

The PMNS mixing matrix $U_\nu\ $ can be parametrized in the standard form
\cite{PDG}
\begin{align} \label{an1}
U_\nu =\left(
\begin{array}{ccc}
c_{12} c_{13} & s_{12} c_{13} & s_{13} e^{-i \delta} \\
-s_{12} c_{23} - c_{12} s_{23} s_{13}  e^{i \delta}
& \quad c_{12} c_{23}  - s_{12} s_{23} s_{13} e^{i \delta} \quad
& s_{23} c_{13} \\
s_{12} s_{23} - c_{12} c_{23} s_{13} e^{i \delta}
& -c_{12} s_{23} - s_{12} c_{23} s_{13} e^{i \delta}
& c_{23} c_{13}
\end{array}\right)\, \cdot P \ ,
\end{align}
where $c_{ij} \equiv \cos \theta_{ij}\ , \ s_{ij} \equiv \sin \theta_{ij}\ $
and $P={\rm diag\ }(1, e^{i \alpha}, e^{i \beta})$; $\delta$ is a Dirac-type
phase (analogous to that of the quark sector) and $\alpha, \beta$ are two
physical phases associated with the Majorana character of neutrinos.

We shall assume for the Dirac neutrino mass spectrum the SO(10)-motivated
hierarchy given in Eq.~(\ref{gf14}), i.e. $M_D \sim M_u\ $. Moreover, since for
the up-quark masses at GUT scale the hierarchy $m_u:m_c:m_t \sim
\epsilon^2:\epsilon:1$ is verified ($\epsilon \simeq 3 \times 10^{-3}$), we
shall write
\begin{align} \label{an2}
d_D = m_t\, \text{diag} (\epsilon^2, \epsilon, 1)\ .
\end{align}

Finally, in order to simplify our analytical discussion, we neglect the
possible misalignment between the charged-lepton and Dirac neutrino mass
matrices. In other words, we assume $V_L \simeq \openone$ in Eqs.~(\ref{gf13})
and (\ref{gf16}). As discussed in Section \ref{sec2} such an approximation is
reasonable in the context of SO(10), where one expects $V_L$ to be of the order
of the CKM matrix. A more refined study which includes this effect will be
given in the next section when we present our full numerical discussion.

Eqs.~(\ref{an1}) and (\ref{an2}) together with the light neutrino mass spectrum
are therefore the only input parameters necessary in our further analysis.

\bigskip

\textbf{Case I: Large mixing (LMA, LOW, VO, Just-So$^2$)}

\bigskip

We consider maximal mixing in the 2-3 sector of the leptonic mixing matrix, i.e
$\theta_{23} = \pi/4$, but keep the solar angle $\theta_{12}$ as a free
parameter in order to account for deviations from maximal mixing in the 1-2
sector. Since our goal is to obtain an upper bound for the baryon asymmetry,
for the moment we can neglect any $CP$-violating phase, i.e. assume $\delta =
\alpha = \beta = 0$, and maximize the asymmetry by simply replacing the
imaginary parts in the lepton asymmetry (\ref{lepto1}) with their corresponding
absolute values. Finally, since the mixing angle $\theta_{13}$ is constrained
by reactor neutrino experiments to be small, $U_{e3} \equiv |\sin \theta_{13}|
\lesssim 0.2$ \cite{CHOOZ}, the PMNS mixing matrix (\ref{an1}) can be
approximately written in the form
\begin{align} \label{an3}
U_{\nu} = \left(\begin{array}{ccc}
c_\odot &\quad s_\odot &\quad U_{e3}\\
-\frac{1}{\sqrt{2}} (s_\odot+c_\odot U_{e3})
&\quad \frac{1}{\sqrt{2}} (c_\odot-s_\odot U_{e3})
&\quad\frac{1}{\sqrt{2}}\\
\frac{1}{\sqrt{2}} (s_\odot-c_\odot U_{e3})  &\quad
-\frac{1}{\sqrt{2}} (c_\odot+s_\odot U_{e3}) &\quad\frac{1}{\sqrt{2}}
\end{array}\right)\ ,
\end{align}
where $s_\odot \equiv \sin \theta_{12},\ c_\odot \equiv \cos \theta_{12}$. The
effective neutrino mass matrix given in terms of the light neutrino masses is
easily obtained from Eq.~(\ref{gf15}).

Now we introduce the symmetric matrix
\begin{align} \label{an4}
M' = \frac{m_t^2 \epsilon^4}{\Delta}d_D^{-1} M_\nu d_D^{-1}\ ,
\quad \Delta=m_1 c_\odot^2+m_2 s_\odot^2 +  m_3 U_{e3}^2\ ,
\end{align}
which can be expressed as
\begin{align} \label{an5}
M'= \left(\begin{array}{ccc}
 1 &\quad p\,\epsilon  &\quad r\,{\epsilon}^2 \\
p\,\epsilon &\quad q\,{\epsilon}^2  & \quad s\,{\epsilon}^3  \\
 r\,{\epsilon}^2 &\quad s\,{\epsilon}^3  &\quad t\,{\epsilon}^4
\end{array}\right)
\end{align}
with the coefficients
\begin{align} \label{an6}
p &= \frac{1}{\sqrt{2} \Delta} [-m_1 (s_\odot+ c_\odot U_{e3}) +m_2 s_\odot
 (c_\odot- s_\odot U_{e3})+m_3 U_{e3}]\ , \nl
q &= \frac{1}{2 \Delta} [m_1 (s_\odot+c_\odot U_{e3})^2 + m_2 (c_\odot- s_\odot
U_{e3})^2
 +m_3]\ , \nl
r &= \frac{1}{\sqrt{2} \Delta} [m_1 c_\odot (s_\odot- c_\odot U_{e3}) - m_2
s_\odot
  (c_\odot + s_\odot U_{e3})+m_3 U_{e3}]\ , \nl
s &= \frac{1}{2 \Delta} [m_1 (c_\odot^2 U_{e3}^2-s_\odot^2) - m_2 (c_\odot^2-
s_\odot^2 U_{e3}^2)
 +m_3]\ , \nl
t &= \frac{1}{2 \Delta} [m_1 (s_\odot- c_\odot U_{e3})^2 + m_2 (c_\odot
+s_\odot U_{e3})^2
 +m_3]\ .
\end{align}

The eigenvalues of $M'$ are easily obtained in leading order of $\epsilon$. We
find
\begin{align} \label{an7}
\lambda_1 \simeq 1 + p^2 \epsilon^2\ ,\ \lambda_2 \simeq (q-p^2)\,
\epsilon^2\ , \ \lambda_3 \simeq \frac{s\, (2 p\, r - s) + q
(t-r^2)-p^2\, t}{q-p^2}\, \epsilon^4\ .
\end{align}
According to Eqs.~(\ref{gf16}) and (\ref{an4}), these eigenvalues are nothing
but the inverse of the right-handed Majorana masses with the proper
normalization coefficient. Thus from Eqs.~(\ref{an4})-(\ref{an7}) we obtain
\begin{align} \label{an8}
M_i = \frac{m_t^2 \epsilon^4}{\Delta} \lambda_i^{-1}\ ,\
i=1,2,3\ .
\end{align}

Next, we can find the unitary matrix $U_R$ which diagonalizes $M'$ and
determines the structure of the Dirac neutrino mass matrix through the
relations (\ref{gf13}) and (\ref{gf16}), i.e.
\begin{align} \label{an9}
M_D = d_D U_R\ .
\end{align}
In leading order of $\epsilon$ we obtain
\begin{align} \label{an10}
U_R \simeq \left(\begin{array}{ccc}
 1-\frac{p^2}{2}\,\epsilon ^2 &\quad p\,\epsilon &\quad
 \frac{p\,s-q\,r}{q-p^2}\,\epsilon^2
 \\
p\,\epsilon &\quad -1+\frac{p^2}{2}\,\epsilon ^2
+\frac{(p\,r-s)^2}{2\,(q-p^2)^2}\,\epsilon^2
&\quad \frac{p\,r-s}{q-p^2}\,\epsilon \\
r\,\epsilon^2 &\quad \frac{p\,r-s}{q-p^2}\,\epsilon &\quad
1-\frac{(p\,r-s)^2}{2\,(q-p^2)^2}\,\epsilon^2
\end{array}\right)\ .
\end{align}

It is now straightforward to estimate the baryon asymmetry. From
Eqs.~(\ref{lepto3}), (\ref{lepto4}), (\ref{an2}) and (\ref{an8})-(\ref{an10})
we find

\begin{align} \label{an11}
Y_B=\frac{3}{32 \pi} \frac{d}{g_\ast}  \frac{p^2 (q-p^2-2 r^2)+2
p\, r s + r^2 (t - r^2)}{1+p^2+r^2} \, \frac{m_t^2}{v^2}\,
\epsilon^4\ .
\end{align}

We can also write down the parameter $\kappa$ which together with the mass
$M_1$ controls the washout effects. Using Eq.~(\ref{lepto9}) we find
\begin{align} \label{an12}
\kappa=\frac{M_P\, \Delta\, (1+p^2+r^2)}{1.7 \times 8 \pi v^2
\sqrt{g_{\ast}}} \ .
\end{align}

As mentioned before we are particularly interested in two limiting cases,
namely, when the light neutrino spectrum is hierarchical or inverted
hierarchical. Below we shall consider each one of these situations.

\bigskip

\textbf{(\,i\,) Hierarchical spectrum}

In this case $m_1 \ll m_2 \simeq \sqrt{\dmsol} \ll m_3 \simeq \sqrt{\dmatm}\ $
and the coefficients in the matrix (\ref{an5}) are approximately given by
\begin{align} \label{an13}
p &\simeq \frac{m_2 s_{2 \odot} +2 m_3\,U_{e3}}{2 \sqrt{2}(m_2 s_\odot^2 +
m_3\,U_{e3}^2)}\ , \quad  q \simeq  s \simeq t
\simeq \frac{m_3}{ 2( m_2 s_\odot^2+m_3\,U_{e3}^2)}\ , \nl
r  &\simeq \frac{-m_2 s_{2 \odot} +2 m_3 U_{e3}}{2 \sqrt{2}\,
(m_2 s_\odot^2 + m_3 U_{e3}^2)}\ ,
\end{align}
with $ s_{2 \odot} \equiv \sin 2 \theta_{12}$.

The right-handed Majorana neutrino masses read as
\begin{align} \label{an14}
M_1 &\simeq \frac{m_t^2 \epsilon^4}{m_2 s_\odot^2+ m_3 U_{e3}^2} \simeq
\frac{m_u^2}{s_\odot^2\sqrt{\dmsol}+ \sqrt{\dmatm}\, U_{e3}^2}\ , \nl
 M_2 &\simeq \frac{2 m_t^2 \epsilon^2}{m_2 m_3 s_\odot^2}\,\left(m_2 s_\odot^2
 + m_3 U_{e3}^2\right)
\simeq \frac{2 m_c^2\,\left(s_\odot^2 \sqrt{\dmsol}+ \sqrt{\dmatm}
U_{e3}^2 \right)}{s_\odot^2 \sqrt{\dmsol \dmatm}} \ , \nl
 M_3 &\simeq \frac{m_t^2 s_\odot^2}{2 m_1}\ .
\end{align}

We notice that the requirement $M_3 \lesssim M_P$ implies a lower bound on the
lightest neutrino mass:
\begin{align} \label{an15}
m_1 \gtrsim \frac{m_t^2 s_\odot^2}{2 M_P} \simeq 2 \times 10^{-7}\ \text{eV}\ ,
\end{align}
for a typical value of $m_t \simeq 100$~GeV at GUT scale and a large solar
mixing angle $\theta_{12} \simeq \pi/4$. Obviously, deviations from maximal
solar mixing will slightly modify this bound.

We can also set an upper bound for the lightest right-handed Majorana mass
$M_1$. Indeed, from the first equation in (\ref{an14}) and taking $U_{e3}=0$ we
find
\begin{align} \label{an16a}
M_1 \lesssim \frac{m_u^2}{s_\odot^2\sqrt{\dmsol}}\ .
\end{align}
With $s_\odot$ and $\dmsol$ given by the lower bounds in the case of the
Just-So$^2$ solar (cf. Table \ref{table1}) we find $M_1 \lesssim 1.3 \times
10^9$~GeV. We notice that this value is consistent with the requirement $M_1 <
T_{RH}$, where $T_{RH}$ is the reheating temperature after inflation, which in
turn is constrained to be below $10^8 - 10^{10}$~GeV from considerations of the
gravitino problem \cite{giudice}.

Moreover, for the mass ratios $M_1/M_i\ ,\ i=2,3$ we have
\begin{align} \label{an17a}
\frac{M_1}{M_2} \simeq \frac{m_u^2 r_\odot s_\odot^2}{ 2 m_c^2 (r_\odot
s_\odot^2+ U_{e3}^2)^2}\ , \quad \frac{M_1}{M_3} \simeq \frac{2 m_u^2
m_1}{m_t^2 s_\odot^2 \sqrt{\dmatm}(r_\odot s_\odot^2 +  U_{e3}^2)} \ll
\frac{M_1}{M_2}\ ,
\end{align}
where
\begin{align} \label{an18}
r_\odot \equiv \left(\frac{\dmsol}{\dmatm}\right)^{1/2}\ .
\end{align}

Let us also write down the approximate expression for the baryon asymmetry:
\begin{align} \label{an16}
Y_B \simeq \frac{3}{32 \pi} \frac{d}{g_\ast} \frac{m_2\, m_3^3\,
s_\odot^2\,U_{e3}^2}{(m_2 s_\odot^2+ m_3 U_{e3}^2)^2\, (m_2^2
s_\odot^2 + m_3^2 U_{e3}^2)}\frac{m_t^2}{v^2}\, \epsilon^4\ ,
\end{align}
i.e.,
\begin{align} \label{an17}
Y_B \simeq \frac{3}{32 \pi} \frac{d}{g_\ast} \frac{m_u^2}{v^2}
\frac{r_\odot\, s_\odot^2\, U_{e3}^2 }{(r_\odot s_\odot^2+
U_{e3}^2)^2 (r_\odot^2 s_\odot^2+ U_{e3}^2)} \ .
\end{align}
As a function of $U_{e3}\ $, $Y_B$ reaches its maximum value when
\begin{align} \label{an19}
U_{e3}^2 = \frac{s_\odot^2\,r_\odot^{3/2}}{\sqrt{2}}\ .
\end{align}
Substituting this value into Eq.~({\ref{an17}) we obtain the upper bound
\begin{align} \label{an20}
Y_B \lesssim \frac{3}{32 \pi} \frac{d}{g_\ast} \frac{m_u^2}{v^2} \,
\frac{1}{r_\odot s_\odot^2} \simeq 9.2 \times
10^{-15}\,\frac{d}{r_\odot s_\odot^2}\ .
\end{align}

It is then clear that the larger the mass squared difference $\dmsol$ is, the
higher is the suppression in the $CP$ asymmetry $\epsilon_{N_1}$. Thus, while
the Just-So$^2$ vacuum oscillation solution is the most favoured in this
framework, the LMA solar solution turns out to be highly disfavoured.

\begin{table*}
\caption{Constraints on neutrino masses and mixing angles coming from global
analyses of the solar, atmospheric and reactor neutrino data
\protect\cite{gonzalez,bahcall}. The numbers in square brackets correspond to
the best-fit values.}
\medskip
\renewcommand{\tabcolsep}{0.4pc}
\begin{tabular}{lccc} \hline \hline \noalign{
\medskip
\begin{center}
Atmospheric and reactor neutrinos
\end{center} \smallskip
} \hline \noalign{\smallskip}
 & $\dmatm~(\text{eV}^2)$ & $\tgatm$ & $U_{e3}$ \\
\noalign{\smallskip} \hline \noalign{\smallskip}
& $(\,1.4 - 6.1 \,)\,[\,3.1\,] \times 10^{-3}$ & (\,0.4 - 3.1\,)\,[\,1.4\,] & $ < 0.2$ \\
\noalign{\medskip} \hline \hline \noalign{
\medskip
\begin{center}
Solar neutrinos
\end{center} \smallskip
} \hline \noalign{\smallskip} \noalign{\smallskip}
 & $\dmsol~(\text{eV}^2)$ & $\tgsol$ & $r_\odot= \left(\dmsol/\dmatm\right)^{1/2}$ \\
\noalign{\smallskip} \hline \noalign{\smallskip} LMA & $(\,2 - 30
\,)\,[\,4.5\,] \times 10^{-5} $
& (\,0.3 - 1\,)\,[\,0.4\,]  & $(\,0.6-4.6\,)\,[\,1.2\,] \times 10^{-1}$ \\
\noalign{\smallskip} SMA & $(\,4-10\,)\,[\,4.7\,] \times
10^{-6}$ & $(\,2-8\,)\,[\,4\,] \times 10^{-4}$ & $(\,2.6-8.4\,)\,[\,3.9\,] \times 10^{-2}$ \\
\noalign{\smallskip} LOW & $(\,0.3-2\,)\,[\,1.0\,] \times
10^{-7}$ & $(\,0.5-1.1\,)\,[\,0.7\,]$
& $(\,0.2-1.2\,)\,[\,0.6\,] \times 10^{-2}$ \\
\noalign{\smallskip} VO & $(\,4-6\,)\,[\,4.6\,] \times 10^{-10}$
& $(\,1.5-4\,)\,[\,2.4\,]$
& $(\,2.6-6.5\,)\,[\,3.8\,] \times 10^{-4}$ \\
\noalign{\smallskip} Just-So$^2$ & $(\,5-8\,)\,[\,5.5\,] \times
10^{-12}$ &
$(\,0.5-2\,)\,[\,0.7\,]$& $(\,2.9-7.6\,)\,[\,4.2\,] \times 10^{-5}$ \\
\noalign{\smallskip} \hline \hline \medskip
\end{tabular}
\label{table1}
\end{table*}

In Table~\ref{table2} we give the upper bounds for the different solar
solutions obtained using the present constraints on neutrino masses and mixing
angles coming from global analyses of the solar, atmospheric and reactor
neutrino data (cf. Table~\ref{table1}). We see that even if one neglects the
washout effects (i.e. if one assumes $d=1$), the LOW and LMA solutions have an
upper bound which is below the lower bound of the observed baryon asymmetry. On
the other hand, the VO and Just-So$^2$ solar solutions have an upper bound
which lies inside the allowed range for the asymmetry.

A more realistic bound can be obtained by including the dilution effects. From
Eq.~(\ref{an12}) we find
\begin{align} \label{an21}
\kappa=\frac{M_P}{1.7 \times 8 \pi v^2 \sqrt{g_{\ast}}} \frac{m_2^2 s_\odot^2+
m_3^2 U_{e3}^2}{m_2 s_\odot^2 +  m_3 U_{e3}^2} \simeq \frac{9.1 \times
10^2}{1\, \text{eV}} \frac{s_\odot^2 \dmsol +  \dmatm U_{e3}^2}{s_\odot^2
\sqrt{\dmsol} +  \sqrt{\dmatm} U_{e3}^2}\ .
\end{align}
Here we can distinguish two different regimes for the parameter $\kappa$
depending on the value of $U_{e3}$:
\begin{align} \label{an22}
\kappa &\simeq 9.1 \times 10^2 \frac{\sqrt{\dmsol}}{\text{1 eV}} \quad
\text{for} \quad U_{e3}^2 \ll r_\odot^2 s_\odot^2 \ll r_\odot s_\odot^2\ , \nl
\kappa &\simeq 9.1 \times 10^2 \frac{\sqrt{\dmatm}}{\text{1 eV}} \quad
\text{for} \quad U_{e3}^2 \gg r_\odot s_\odot^2 \gg r_\odot^2 s_\odot^2\ .
\end{align}

Thus, for large values of $U_{e3}$ the parameter $\kappa$ is controlled by the
mass squared difference of the atmospheric neutrinos and for all the large
mixing solar solutions it lies in the range $35 \lesssim \kappa \lesssim 70$.
However, to establish an upper bound on the baryon asymmetry we use the value
of $U_{e3}$ in Eq.~(\ref{an19}) which maximizes $Y_B$. Substituting it into
Eq.~(\ref{an21}) we have
\begin{align} \label{an23}
\kappa \simeq 6.4 \times 10^2 \frac{(\dmatm\, \dmsol)^{1/4}}{\text{eV}}\ ,
\end{align}
and according to Eq.~(\ref{an14}), the mass of the lightest right-handed
Majorana neutrino is approximately given by
\begin{align} \label{an24}
M_1 \simeq \frac{m_u^2}{s_\odot^2 \sqrt{\dmsol}}\ .
\end{align}
Using these expressions to calculate the dilution factor $d$ we obtain the
upper bounds given in Table \ref{table2}. These bounds are of course more
stringent than the ones previously obtained neglecting the washout effects.

\begin{table*}
\caption{Upper bounds for the clean ($d=1$) and net ($d \neq 1$) baryon
asymmetries in the case of hierarchical neutrinos. The bounds are obtained from
Eq.~(\ref{an20}), using the solar, atmospheric and reactor neutrino data of
Table~\ref{table1}.}
\medskip
\begin{center}
\begin{tabular}{lccccc}
\hline \hline  \noalign{\smallskip}
Solution & LMA & LOW & VO &Just-So$^2$ & SMA \\
\noalign{\smallskip} \hline \noalign{\smallskip} $Y_B$ (clean)  & $3.1 \times
10^{-13} $ & $8.7 \times 10^{-12} $
&$4.3 \times 10^{-11} $ & $3.2 \times 10^{-10} $ & $3.3 \times 10^{-10} $\\
\noalign{\smallskip} $Y_B$ (net)  & $1.4 \times 10^{-14}$ & $5.1 \times
10^{-13}$ & $2.5 \times 10^{-12}$ & $2.0 \times 10^{-11} $
& $1.7 \times 10^{-11}$\\
 \hline \hline \\
\end{tabular}
 \label{table2}
\end{center}
\end{table*}

In Fig.~\ref{fig1} we plot the asymmetry $Y_B$ as a function of $U_{e3}$ for
the mass of the lightest neutrino $m_1 = 10^{-6}$~eV. Fig.~\ref{fig1}a
corresponds to the clean asymmetry, i.e. neglecting the dilution effects, while
in Fig.~\ref{fig1}b the net baryon asymmetry (after including the washout
effects) is plotted. The parameter $\kappa$ as defined by Eq.~(\ref{lepto9}) is
given in Fig.~\ref{fig1}c. We notice that for very small values of $U_{e3}$
this parameter is proportional to the mass squared difference of the solar
neutrinos, $\kappa \propto (\dmsol)^{1/2}\,$, and as $U_{e3}$ increases, it
tends to a common value for all the solar solutions, which is determined by the
mass squared difference of the atmospheric neutrinos, i.e. $\kappa \propto
(\dmatm)^{1/2}$ (see Eqs.~(\ref{an22})). A similar behaviour is observed for
the lightest Majorana neutrino mass $M_1$ as can be seen from Fig.~\ref{fig1}d.
In the latter case, $M_1 \propto (\dmsol)^{-1/2}\,$ for small values of
$U_{e3}$ and $M_1 \propto (\dmatm)^{-1/2}\,$ for large values of $U_{e3}$ (see
also Eqs.~(\ref{an14})).

\begin{figure*}
$$\includegraphics[width=14cm]{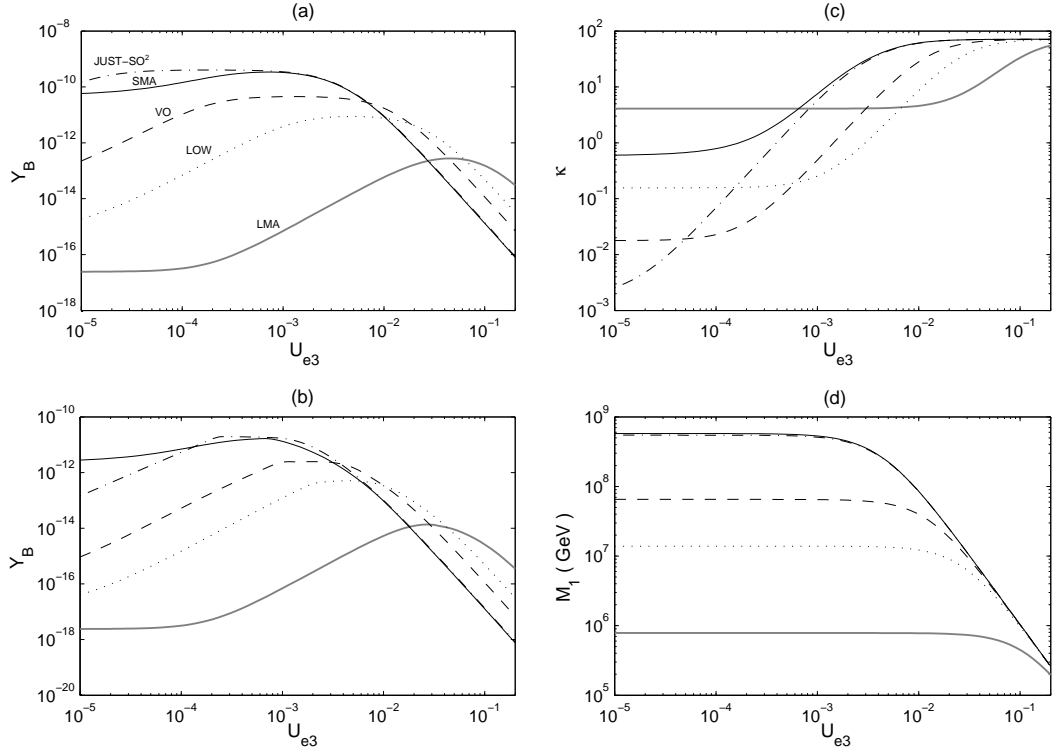}$$
\caption{The baryon asymmetry $Y_B$ as a function of $U_{e3}$ for the different
solar solutions. Fig.~(a) corresponds to the clean asymmetry, while in Fig.~(b)
the net baryon asymmetry, which includes the washout effects, is given. The
parameter $\kappa$ defined in Eq.~(\ref{lepto9}) and the mass of the lightest
right-handed Majorana neutrino $M_1$ are plotted in Figs.~(c) and (d),
respectively. The curves are given for the mass of the lightest neutrino $m_1 =
10^{-6}$~eV. The values for the neutrino oscillation parameters are taken from
Table~\ref{table1}.} \label{fig1}
\end{figure*}

At this point one may wonder whether the inclusion of a misalignment between
the charged-lepton and Dirac neutrino mass matrices could change our
conclusions. Since in SO(10)-motivated scenarios such a misalignment is
expected to be proportional to the CKM quark mixing matrix, let us then assume
for the unitary matrix $V_L$ in Eq.~(\ref{gf7}) the following CKM-type real
matrix
\begin{align} \label{wolf}
V_{L} \simeq \left(\begin{array}{ccc}
1-\lambda^2/2 &\quad \lambda &\quad 0\\
-\lambda &\quad 1-\lambda^2/2 &\quad A \lambda^2\\
0 &\quad -A \lambda^2 &\quad1
\end{array}\right) \ ,
\end{align}
with $A \simeq 1$ and $\lambda$ of the order of the Cabibbo angle, i.e.
$\lambda \lesssim 0.2$.

Following the same analytic approach we can obtain upper bounds for the baryon
asymmetry. Since the relevant formulae are more cumbersome in this case, it is
more illustrative to present the results in a simple plot, similar to that of
Fig.~\ref{fig1}. The results are presented in Fig.~\ref{fig2}. From this figure
it is seen that the upper bounds previously found for $Y_B$ are essentially
unaltered and that our previous conclusions remain valid in this case. To
maximize the effect we have taken the Dirac phase $\delta$ in the PMNS neutrino
mixing matrix to be $\pi$, as suggested from our full numerical study (see next
section). The resonance behaviour in the curves at $U_{e3} \simeq 0.1$ is
associated with our particular choice for the values of $\delta$ and $\lambda$.

\begin{figure*}
$$\includegraphics[width=14cm]{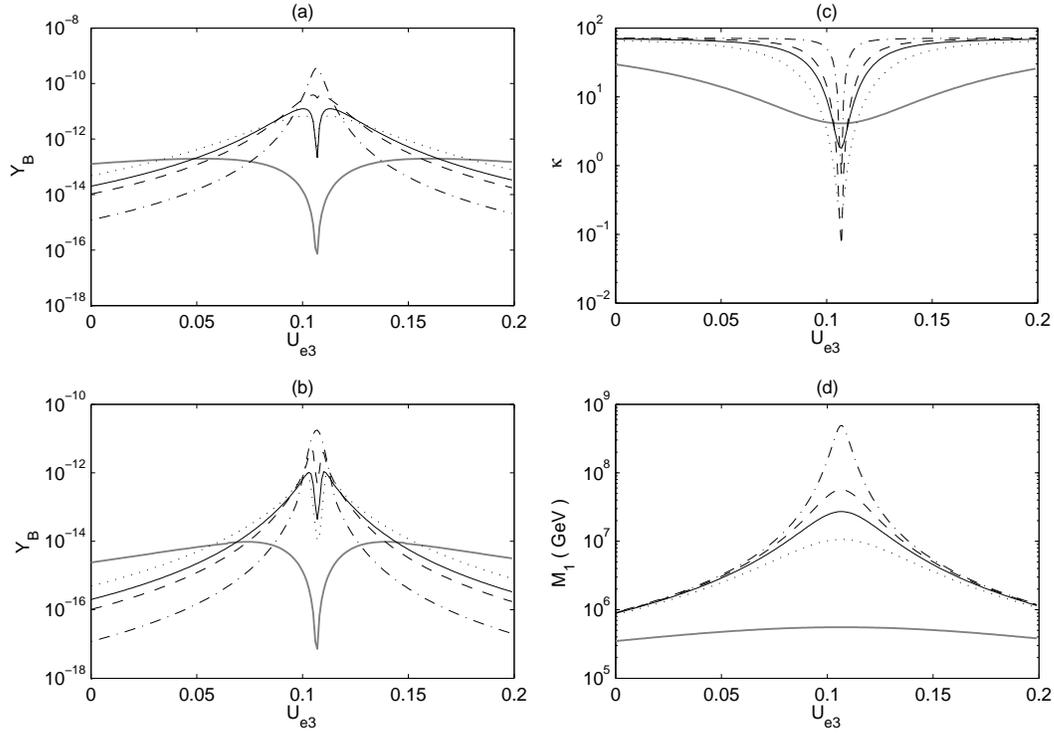}$$
\caption{The same plots as in Fig.~\ref{fig1} but for $\lambda = 0.15$ and a
Dirac phase $\delta = \pi$.} \label{fig2}
\end{figure*}

\textbf{(\,ii\,) Inverted-hierarchical spectrum}

In this case $m_3 \ll m_1 \simeq  m_2 \simeq \sqrt{\dmatm}\ $ and the
coefficients in the matrix (\ref{an5}) are approximately given by the
expressions:
\begin{align} \label{an25}
p &\simeq \frac{(m_2-m_1) s_{2 \odot}}{2 \sqrt{2}(m_1 c_\odot^2 +m_2 s_\odot^2)}
-\frac{U_{e3}}{\sqrt{2}}\ , \nl
q &\simeq \frac{m_1 s_\odot^2 +m_2 c_\odot^2}{2(m_1 c_\odot^2 +m_2 s_\odot^2)}
-\frac{(m_2-m_1) s_{2 \odot}U_{e3}}{4(m_1 c_\odot^2 +m_2 s_\odot^2)}\ , \nl
r &\simeq -\frac{(m_2-m_1) s_{2 \odot}}{2 \sqrt{2}(m_1 c_\odot^2 +m_2 s_\odot^2)}
-\frac{U_{e3}}{\sqrt{2}}\ , \nl
s &\simeq -\frac{m_1 s_\odot^2 +m_2 c_\odot^2}{2(m_1 c_\odot^2 +m_2 s_\odot^2)}
+\frac{1}{2} U_{e3}^2\ ,\nl
t &\simeq \frac{m_1 s_\odot^2 +m_2 c_\odot^2}{2(m_1 c_\odot^2 +m_2 s_\odot^2)}
+\frac{(m_2-m_1) s_{2 \odot} U_{e3}}{4(m_1 c_\odot^2 +m_2 s_\odot^2)}\ ,
\end{align}
with
\begin{align} \label{an25a}
\frac{m_2-m_1}{m_1 c_\odot^2 +m_2 s_\odot^2} \simeq \frac{1}{2}
\frac{\dmsol}{\dmatm} = \frac{r_\odot^2}{2} \ .
\end{align}
The right-handed Majorana neutrino masses are given in this limit by
\begin{align} \label{an26}
M_1 &\simeq \frac{m_t^2 \epsilon^4}{m_1 c_\odot^2 +m_2 s_\odot^2} \simeq
\frac{m_u^2}{\sqrt{\dmatm}}\ , \nl
 M_2 &\simeq 2 m_t^2
\epsilon^2\,\left(\frac{c_\odot^2}{m_2}+\frac{s_\odot^2}{m_1}\right) \simeq
\frac{2 m_c^2}{\sqrt{\dmatm}}\ , \nl
 M_3 &\simeq \frac{m_t^2}{4 m_3} \frac{m_1+m_2}{m_1 c_\odot^2 +m_2 s_\odot^2}
 \simeq \frac{m_t^2}{2 m_3}\ .
\end{align}

Once again requiring $M_3 \lesssim M_P$ and taking $m_t \simeq 100$~GeV at GUT
scale, we find the following lower bound on the lightest neutrino $m_3\ $,
\begin{align} \label{an27}
m_3 \gtrsim \frac{m_t^2}{2 M_P} \simeq 4 \times 10^{-7}\
\text{eV}\ .
\end{align}

The generated baryon asymmetry can be approximated in this case by the
following expression
\begin{align} \label{an28}
Y_B \simeq \frac{3}{128 \pi} \frac{d}{g_\ast}  \frac{m_1 m_2
(m_1-m_2)^2 s_{2 \odot}^2 + 2 m_3 (m_1^3+m_2^3) U_{e3}^2}{(m_1^2
c_\odot^2 +m_2^2 s_\odot^2) (m_1 c_\odot^2 +m_2 s_\odot^2)^2}\,
\frac{m_t^2}{v^2}\, \epsilon^4\ ,
\end{align}
or equivalently,
\begin{align} \label{an29}
Y_B \simeq \frac{3}{512 \pi} \frac{d}{g_\ast} \frac{m_u^2}{v^2} \left[
s_{2 \odot}^2 r_\odot^4 + \frac{16 m_3 U_{e3}^2}{\sqrt{\dmatm}} \right]\ .
\end{align}
Taking for instance $m_3 \simeq \sqrt{\dmsol} \ll m_{1,2} \simeq \sqrt{\dmatm}$
and the maximum value of $U_{e3} \simeq 0.2$ it is easy to see that the
dominant contribution to the asymmetry comes from the second term in the square
brackets. Neglecting the washout effects ($d=1$) we find the following upper
bound:
\begin{align} \label{an30}
Y_B \lesssim \frac{3}{32 \pi} \frac{d}{g_\ast} \frac{m_u^2}{v^2}
U_{e3}^2 r_\odot \simeq 3.7 \times 10^{-16} r_\odot \ .
\end{align}

Thus we conclude that the baryon asymmetry is highly suppressed for all the
large mixing solar solutions and for an inverted hierarchical neutrino
spectrum, even without taking into account the washout effects. It is however
worthwhile to have an idea of these effects. The parameter $\kappa$ reads as
\begin{align} \label{an31}
\kappa= \frac{M_P}{1.7 \times 16 \pi v^2 \sqrt{g_{\ast}}}
\frac{m_1^2+m_2^2}{m_1 c_\odot^2+m_2 s_\odot^2}
\simeq \frac{M_P \sqrt{\dmatm}}{1.7 \times 8 \pi v^2 \sqrt{g_{\ast}}} \ .
\end{align}
Therefore,
\begin{align} \label{an32}
\kappa \simeq 9.1 \times 10^2 \left[ \frac{\sqrt{\dmatm}}{\text{eV}}\right]
\simeq 50\ .
\end{align}
Since from Eq.~(\ref{an26}) we have $M_1 \simeq 1.8 \times 10^4$~GeV,
then from Eqs.(\ref{lepto10}) we find $d \simeq 10^{-2}$.

\bigskip

\textbf{Case II: Small mixing (SMA)}

\bigskip

Let us now consider the small mixing solar solution. Most of the relevant
formulas can be easily obtained from the previously discussed large mixing case
by just letting $c_\odot \rightarrow 1$ and assuming $s_\odot \ll 1$. We must
however proceed with care since in this case there are two small parameters
competing against each other, namely, the small solar mixing angle
$\theta_{12}$ and $U_{e3}$. In this case the leptonic mixing matrix $U_\nu$ can
be approximated as

\begin{align} \label{an33}
U_{\nu}= \left(\begin{array}{ccc}
1 &\quad s_\odot &\quad U_{e3}\\
-\frac{1}{\sqrt{2}}(s_\odot+U_{e3}) &\quad
\frac{1}{\sqrt{2}}(1-s_\odot U_{e3}) &\quad \frac{1}{\sqrt{2}}\\
\frac{1}{\sqrt{2}} (s_\odot-U_{e3})
&\quad-\frac{1}{\sqrt{2}}(1+s_\odot U_{e3}) &\quad\frac{1}{\sqrt{2}}
\end{array}\right)\ .
\end{align}

Next we follow the same steps performed in the large mixing case. We first
define the new matrix
\begin{align} \label{an34}
M' = \frac{m_t^2 \epsilon^4}{\Delta}d_D^{-1} M_\nu d_D^{-1}\ ,
\quad \Delta=m_1+m_2 s_\odot^2+ m_3 U_{e3}^2\ ,
\end{align}
which can be written in the form (\ref{an5}) with the coefficients $p, q, r, s,
t$ as given in (\ref{an6}). The eigenvalues and eigenvectors of $M'$ are given
by the same expressions in Eqs.~(\ref{an7}) and (\ref{an10}), respectively. The
right-handed Majorana masses read as in Eq.~(\ref{an8}). Finally, the baryon
asymmetry will be given by Eq.~(\ref{an11}) with the parameter $\kappa$
defined in Eq.~(\ref{an12}).

Let us now consider the limiting cases of hierarchical and
inverted-hierarchical neutrino mass spectrum.

\bigskip

\textbf{(\,i\,) Hierarchical spectrum}

In the limit $m_1 \ll m_2 \ll m_3$ we find
\begin{align} \label{an38}
p &\simeq \frac{m_2 s_\odot+m_3 U_{e3}}{\sqrt{2}\,(m_1+m_2 s_\odot^2+ m_3
U_{e3}^2)}  \ , \quad q \simeq s \simeq t \simeq \frac{m_3}{2\,(m_1+m_2
s_\odot^2+ m_3 U_{e3}^2)}\ , \nl r &\simeq \frac{-m_2 s_\odot+m_3
U_{e3}}{\sqrt{2}\, (m_1+m_2 s_\odot^2+ m_3 U_{e3}^2)} \ .
\end{align}

Moreover,
\begin{align} \label{an39}
M_1 &\simeq \frac{m_t^2 \epsilon^4}{m_1+m_2 s_\odot^2+m_3 U_{e3}^2} \simeq
\frac{m_u^2}{m_1+\sqrt{\dmsol}\, s_\odot^2 + \sqrt{\dmatm}\, U_{e3}^2}\ ,\nl \nl
M_2 &\simeq \frac{2m_t^2 \epsilon^2}{m_3}
\frac{m_1+m_2 s_\odot^2+m_3 U_{e3}^2}{m_1+m_2 (s_\odot - U_{e3})^2}
\simeq \frac{2m_c^2}{\sqrt{\dmatm}}
\frac{m_1+\sqrt{\dmsol}\, s_\odot^2+\sqrt{\dmatm}\, U_{e3}^2}{m_1+\sqrt{\dmsol}\,
(s_\odot - U_{e3})^2}\ ,\nl \nl
M_3 &\simeq \frac{m_t^2}{2} \left(\frac{1}{m_2}+\frac{(s_\odot - U_{e3})^2}{m_1}
\right) \simeq \frac{m_t^2}{2} \left(\frac{1}{\sqrt{\dmsol}}+
\frac{(s_\odot - U_{e3})^2}{m_1} \right)\ .
\end{align}

From the requirement that $M_3$ be smaller than the Planck mass we find the
following lower bound for the mass of the lightest neutrino:
\begin{align} \label{an40}
m_1 \gtrsim \frac{m_t^2 \sqrt{\dmsol} (s_\odot-U_{e3})^2}{2 M_P \sqrt{\dmsol} -m_t^2}
\simeq 4.1 \times 10^{-7}\,(s_\odot-U_{e3})^2~\text{eV}\ .
\end{align}
Using for instance the minimum value $s_\odot = 0.014$ and taking $U_{e3} = 0$
one has $m_1 \gtrsim 8 \times 10^{-11}$~eV. The above bound is of course
sensitive to the value of $U_{e3}$ and it gets weaker as $U_{e3}$ approaches
$s_\odot$.

Let us now estimate the baryon asymmetry. From Eqs.~(\ref{an11}) and
(\ref{an38}) we find
\begin{align} \label{an41}
Y_B &\simeq \frac{3}{32 \pi} \frac{d}{g_\ast} \frac{m_t^2}{v^2}
\frac{m_3^3\, (m_1+m_2 s_\odot^2)\, U_{e3}^2\,\epsilon^4}{(m_1+m_2 s_\odot^2+
m_3 U_{e3}^2)^2 (m_1^2+m_2^2 s_\odot^2+m_3^2 U_{e3}^2)} \nl
&\simeq \frac{3}{32 \pi} \frac{d}{g_\ast} \frac{m_u^2}{v^2} \frac{(\dmatm)^{3/2}
(m_1+s_\odot^2\sqrt{\dmsol} ) U_{e3}^2}{(m_1+s_\odot^2 \sqrt{\dmsol}
+ \sqrt{\dmatm} U_{e3}^2)^2 (m_1^2+s_\odot^2\dmsol  + \dmatm U_{e3}^2)}\ .\nl
\end{align}

We can also compute the parameter $\kappa$ relevant for the dilution effects.
From the definition (\ref{an12}) we find
\begin{align} \label{an42}
\kappa = \frac{M_P}{1.7 \times 8 \pi v^2  \sqrt{g_{\ast}}}
\frac{m_1^2+m_2^2 s_\odot^2+m_3^2 U_{e3}^2}{m_1+m_2 s_\odot^2+m_3 U_{e3}^2}\ .
\end{align}

As a function of $U_{e3}$ the expression (\ref{an41}) reaches its maximum when
\begin{align} \label{an43}
U_{e3}^2 = \frac{(m_1+s_\odot^2\sqrt{\dmsol} )^{1/2}\,
(m_1^2+s_\odot^2\dmsol )^{1/2}}{\sqrt{2} (\dmatm)^{3/4}}\ .
\end{align}
For $m_1 \gg s_\odot \sqrt{\dmsol}  \gg s_\odot^2\,\sqrt{\dmsol} $ we have
\begin{align} \label{an44}
U_{e3}^2 \simeq \frac{1}{\sqrt{2}}
\left(\frac{m_1}{\sqrt{\dmatm}}\right)^{3/2}\ ,
\end{align}
and therefore
\begin{align} \label{an45}
Y_B \lesssim \frac{3}{32 \pi} \frac{d}{g_\ast} \frac{m_u^2}{v^2}
\frac{\sqrt{\dmatm}}{m_1} \simeq 7.2 \times 10^{-16}\, d
\left[\frac{\text{eV}}{m_1} \right]\ .
\end{align}
For the parameter $\kappa$ we have in turn
\begin{align} \label{an46}
\kappa \simeq \frac{6.4 \times 10^2}{1~\text{eV}} m_1^{1/2}\,(\dmatm)^{1/4}
\simeq 1.8 \times 10^2 \sqrt{\frac{m_1}{\text{eV}}}\ .
\end{align}
Finally, for the lightest right-handed Majorana mass we obtain
\begin{align} \label{an47}
M_1 \simeq \frac{m_u^2}{m_1} \simeq 10^3\, \text{GeV}\,
\left[\frac{\text{eV}}{m_1}\right] \ .
\end{align}
It remains to estimate the dilution factor $d$. Assuming e.g. $m_1 =
10^{-4}$~eV, so that the relation $m_1 \gg s_\odot \sqrt{\dmsol}$ is satisfied,
Eqs.~(\ref{an46}) and (\ref{an47}) imply $\kappa \simeq 1.8, M_1 = 10^7$~GeV,
and therefore $d \simeq 0.14$. Thus, from Eq.~(\ref{an45}) we obtain the upper
bound
 \begin{align} \label{an48}
Y_B \lesssim 10^{-12}\ ,
\end{align}
which is one order of magnitude below the required value for the observed
asymmetry.

On the other hand, if $m_1 \ll s_\odot^2 \sqrt{\dmsol}  \ll
s_\odot\,\sqrt{\dmsol} $ then the maximum value in Eq.~({\ref{an41}) is
obtained when
\begin{align} \label{an49}
U_{e3}^2 = \frac{s_\odot^2\,r_\odot^{3/2}}{\sqrt{2}}\ .
\end{align}
In this case, the expression for the upper bound of the baryon asymmetry is the
same as the one previously found for the large mixing solar solutions (see
Eq.~(\ref{an20})):
\begin{align} \label{an50}
Y_B \lesssim \frac{3}{32 \pi} \frac{d}{g_\ast} \frac{m_u^2}{v^2}
\frac{1}{r_\odot s_\odot^2} \simeq 1.8 \times 10^{-9}\, d\ .
\end{align}

The dilution factor turns out to be crucial in obtaining a reliable estimate
for $Y_B$. In this limiting case, the parameter $\kappa$ is also given by the
same expression of the large mixing solutions, i.e. by Eq.~(\ref{an23}). We
find $\kappa \simeq 8$. For the lightest Majorana mass we have in turn,
\begin{align} \label{an51}
M_1 \simeq \frac{m_u^2}{s_\odot^2\,\sqrt{\dmsol}}
\simeq  2.5 \times 10^9~\text{GeV}\ ,
\end{align}
and the mass ratios $M_1/M_2$ and $M_1/M_3$ are given by the approximate
expressions:
\begin{align} \label{an53}
\frac{M_1}{M_2} \simeq \frac{1}{2} \left(\frac{m_u}{m_c}\right)^2 \frac{1}{r_\odot
s_\odot^2}\ , \quad \frac{M_1}{M_3} \simeq 2 \left(\frac{m_u}{m_t}\right)^2
\frac{m_1}{s_\odot^4 \sqrt{\dmsol}}\ .
\end{align}

The above values for $\kappa$ and $M_1$ imply then a dilution factor $d \simeq
8 \times 10^{-3}$. Substituting this value into Eq.~(\ref{an50}) we conclude
that for the SMA solution
\begin{align} \label{an52}
Y_B \lesssim 1.4 \times 10^{-11}\ ,
\end{align}
a bound which is within the allowed range for the baryon asymmetry.

\bigskip

\textbf{(\,ii\,) Inverted-hierarchical spectrum}

Let us now consider an inverted-hierarchical neutrino spectrum, i.e. $m_1
\simeq m_2 \simeq \sqrt{\dmatm} \gg m_3$. In this case the coefficients of the
matrix (\ref{an5}) can be obtained from Eqs.~(\ref{an25}) by setting $c_\odot =
1$ and assuming $s_\odot \ll 1$. We find
\begin{align} \label{an54}
p &\simeq \frac{(m_2-m_1) s_\odot}{\sqrt{2}m_1}
-\frac{U_{e3}}{\sqrt{2}} \simeq \frac{ s_\odot r_\odot^2}{2 \sqrt{2}}
-\frac{U_{e3}}{\sqrt{2}}\ , \quad
q \simeq -s \simeq t \simeq \frac{m_2}{2m_1} \simeq \frac{1}{2}\ , \nl
r &\simeq -\frac{(m_2-m_1) s_\odot}{\sqrt{2}m_1}
-\frac{U_{e3}}{\sqrt{2}} \simeq  -\frac{s_\odot r_\odot^2}{2 \sqrt{2}}
-\frac{U_{e3}}{\sqrt{2}} \ .
\end{align}

The right-handed Majorana neutrino masses will be given by the same expressions
in Eqs.~(\ref{an26}) and, in particular, the lower bound (\ref{an27}) for the
lightest neutrino $m_3$ should be verified as well. Finally, the approximated
expression for the baryon asymmetry is obtained from Eq.~(\ref{an29}):
\begin{align} \label{an55}
Y_B \simeq \frac{3}{32 \pi} \frac{d}{g_\ast} \frac{m_u^2}{v^2}
\frac{m_3 U_{e3}^2 }{\sqrt{\dmatm}} \lesssim 3.7 \times 10^{-16} r_\odot \lesssim
3.1 \times 10^{-17} \ ,
\end{align}
for $d=1, m_3 \simeq \sqrt{\dmsol}$ and $U_{e3} =0.2$. Thus we conclude that
for the SMA solution and an inverted-hierarchical neutrino spectrum the
generated baryon asymmetry is highly suppressed.

\subsection{Numerical Analysis}
\label{sec3b}

From the simple analysis performed in the last section we have concluded that
in the SM framework with the simplest SO(10)-motivated hierarchy (\ref{gf14})
for the Dirac neutrino mass spectrum, only the SMA and Just-So$^2$ solutions of
the solar neutrino problem are compatible with the required value for the
baryon asymmetry of the Universe. In this section we will perform a full
numerical computation of the baryon asymmetry including not only all the
$CP$-violating phases which appear in the PMNS leptonic mixing matrix $U_\nu$
defined in Eq.~(\ref{an1}), but also the possible misalignment which may result
from the process of diagonalization of the charged-lepton and Dirac neutrino
mass matrices. This misalignment is characterized by the matrix $V_L$ in
Eq.~(\ref{gf13}), for which we shall assume the CKM structure of
Eq.~(\ref{wolf}).

We then proceed as follows. The PMNS mixing angles as well as the mass squared
differences $\dmsol$ and $\dmatm$ are allowed to vary in the ranges indicated
in Table~\ref{table1}. First we randomly fix a set of values for the PMNS
mixing angles and the $CP$-violating phases $\alpha$, $\beta$ and $\delta$. The
latter are randomly chosen in the interval from 0 to $2\pi$. In order to
compute the effective neutrino mass matrix we take as an input the mass of the
lightest neutrino, $m_1$, which varies from $10^{-10}$ to 1 eV for the SMA
solution and from $10^{-7}$ to 1 eV for all the large mixing
solutions\footnote{The lower bounds for the mass of the lightest neutrino are
taken in agreement with our previous analytic estimates, so that the
requirement $M_3 \lesssim M_P$ is always satisfied.}. This means that unlike
the analytic study, the full numerical analysis includes also the case of
almost degenerate neutrinos. This is possible because we are taking into
consideration all the phases that may be present and, consequently, possible
cancellations. Assuming $m_1 < m_2 < m_3\ $, the values of $m_2$ and $m_3$ are
determined from the expressions $m_2 = \sqrt{m_1^2+
\dmsol}\,,\,m_3=\sqrt{m_1^2+\dmsol+\dmatm}\ $. The inverse hierarchical case is
not discussed here since it leads to a very suppressed baryon asymmetry as
already shown in the analytic discussion. The effective neutrino mass matrix
$M_\nu$ in Eq.~(\ref{gf15}) is computed, and also the matrix $M$ in
Eq.~(\ref{gf16}) with the up-quark masses given at the GUT scale: $m_u \simeq
1$~MeV, $m_c \simeq 0.3$~GeV and $m_t \simeq 100$~GeV \cite{koide} and the
matrix $V_L$ in Eq.(\ref{wolf}) with $\lambda$ randomly chosen in the interval
$0 \leq \lambda \leq 0.22$. The matrix $M M^\dagger$ is then diagonalized to
obtain $U_R$ and the right-handed neutrino masses $M_i$. In the basis where
$M_\ell$ and $M_R$ are diagonal the Dirac neutrino mass matrix $M_D$ is
determined by Eq.~(\ref{gf13}). Finally we compute the baryon asymmetry from
Eqs.(\ref{lepto1}), (\ref{lepto2}), (\ref{lepto5}) and (\ref{lepto7}), where
the dilution factor $d$ is determined using a combined fit from the
approximations given in Eqs.~(\ref{lepto8a}) and (\ref{lepto10}).

\begin{figure*}
$$\includegraphics[width=14cm]{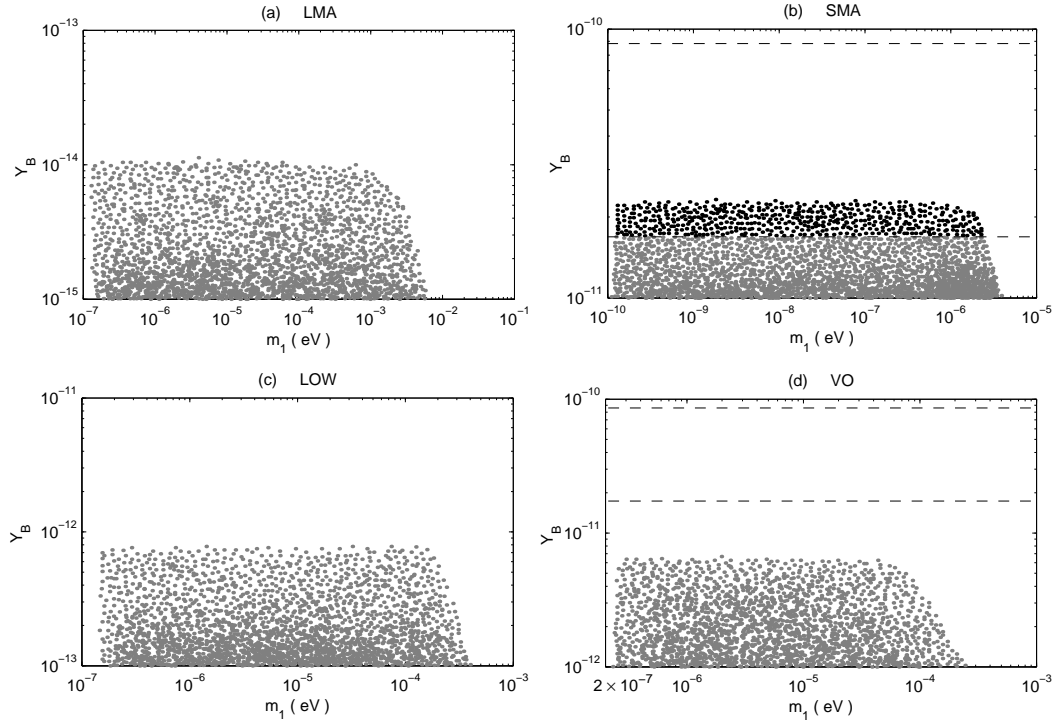}$$
\caption{The baryon asymmetry as a function of the mass $m_1$ of the lightest
neutrino for the LMA, SMA, LOW and VO solar solutions. The black-dotted area
corresponds to sets of input parameters which yield an asymmetry within the
allowed range given in Eq.~(\ref{intro2}) and delimited by the dashed lines.}
\label{fig3}
\end{figure*}

\begin{figure*}
$$\includegraphics[width=14cm]{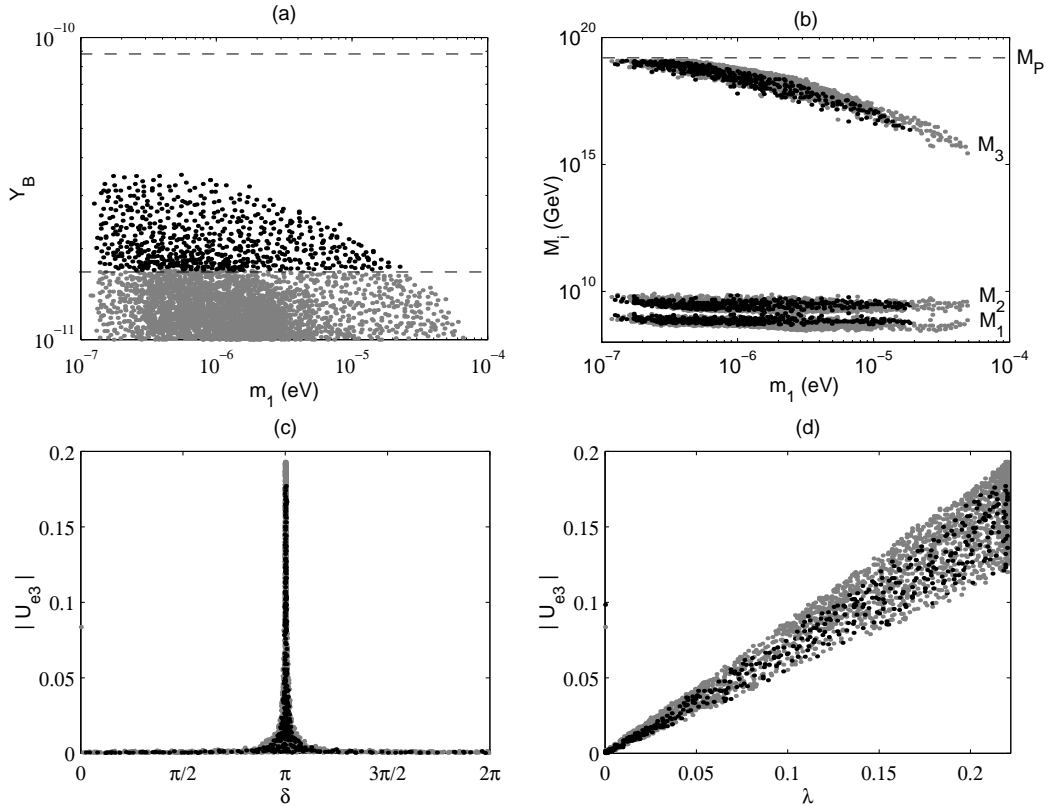}$$
\caption{In Fig.~(a), the baryon asymmetry  as a function of $m_1$ in the case
of the Just-So$^2$ solar solution. The black-dotted area corresponds to sets of
input parameters which give an asymmetry within the allowed range (dashed
lines) as given in Eq.~(\ref{intro2}). In Fig.~(b), the right-handed Majorana
neutrino masses are plotted as functions of $m_1$. From Fig.~(c) we conclude
that for relatively large values of $U_{e3}$ the Dirac phase $\delta$ should be
close to $\pi$ to produce a baryon asymmetry in the allowed range. The allowed
region in the plane $(\lambda, U_{e3})$ is shown in Fig.~(d). The gray-dotted
area is obtained by slightly relaxing the lower bound on the baryon asymmetry,
$Y_B \gtrsim 10^{-11}$. } \label{fig4}
\end{figure*}

In Figs.~\ref{fig3} we present the results for the baryon asymmetry $Y_B$ as a
function of $m_1$ for the LMA, SMA, LOW and VO solutions of the solar neutrino
problem. The first immediate conclusion which can be drawn from these plots is
that among these four solutions only the SMA solution is compatible with the
experimental range for $Y_B$ with a mass for the lightest neutrino in the range
from $2 \times 10^{-10}$ to $4 \times 10^{-6}$ eV. Moreover, the values of
$U_{e3}$ should be small in this case, $|U_{e3}| \lesssim 10^{-2}$. By
comparing the results plotted in Figs.~\ref{fig3} with the upper bounds for
$Y_B$ obtained in the last section (see Table~\ref{table2}), we find the
analytic bounds in very good agreement with the exact numerical ones.

Finally, in Figs.~\ref{fig4} we present the results for the only large mixing
solar solution that produces the required $Y_B$, namely, the Just-So$^2$ vacuum
oscillation solar solution. The asymmetry as a function of the lightest
neutrino mass $m_1$ is plotted in Fig.~\ref{fig4}a. The black-dotted area
corresponds to sets of input parameters which give an asymmetry within the
allowed range delimited by the dashed lines (see also Eq.~(\ref{intro2})). In
Fig.~\ref{fig4}b, the right-handed Majorana neutrino masses are plotted as
functions of $m_1$. As before, the lower bound on $m_1$ is determined by
requiring the mass of the heaviest Majorana neutrino $M_3$ to be below the
Planck scale $M_P$. From Fig.~\ref{fig4}c we conclude that for $|U_{e3}|
\gtrsim 10^{-2}$ the Dirac phase $\delta$ should be close to $\pi$ to produce a
baryon asymmetry in the allowed range. This in turn corresponds to larger
values of the misalignment parameter $\lambda$ in the matrix (\ref{wolf}) as
can be seen from Fig.~\ref{fig4}d.

From Figs.~\ref{fig3} and \ref{fig4} it is also clear that an acceptable baryon
asymmetry requires very small values of the mass of the lightest neutrino, $m_1
\ll 1$~eV. This in particular implies that, in the present framework, an almost
degenerate neutrino spectrum is excluded.

\begin{figure*}
$$\includegraphics[width=14cm]{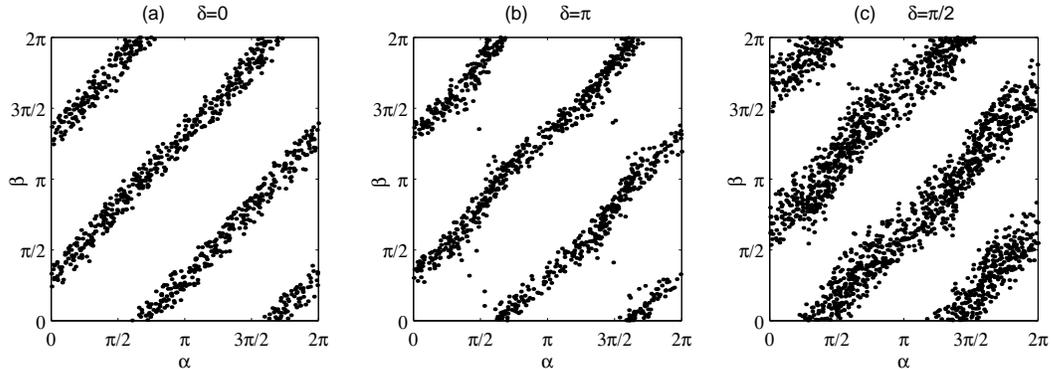}$$
\caption{The $CP$-violating Majorana phases $\alpha$ and $\beta$ in the cases
of no Dirac-type $CP$ violation (Figs. (a) and (b)) and maximal Dirac-type $CP$
violation (Fig. (c)) for the Just-So$^2$ vacuum oscillation solar solution. The
black-dotted area in the plots refers to the values which lead to an acceptable
baryon asymmetry $Y_B$.} \label{fig5}
\end{figure*}

To establish a relationship between $CP$ violation at high and low energies is
not an easy task in the sense that a general analytic treatment with the full
set of parameters would be difficult to perform. Nevertheless, one can
investigate whether there is any correlation between the $CP$-violating phases
$\alpha$, $\beta$ and $\delta$ in a plausible leptogenesis scenario. In
Fig.~\ref{fig5} we present the results of a random analysis in the cases of no
Dirac-type $CP$ violation ($\delta=0,\pi$) and maximal Dirac-type $CP$
violation ($\delta=\pi/2$) for the Just-So$^2$ vacuum oscillation solar
solution, which is the only large mixing solar solution compatible with
leptogenesis in the simplest SO(10)-motivated framework considered here. The
black-dotted area in the plots refers to the values of $\alpha$ and $\beta$
which lead to an acceptable baryon asymmetry $Y_B$. It is seen from
Figs.~\ref{fig5}a and \ref{fig5}b that one can obtain the right magnitude for
$Y_B$ with a single nonvanishing Majorana phase. This means that the possible
absence of $CP$ violation effects in neutrino oscillations in future
experiments may not \emph{a priori} discard leptogenesis as the  mechanism
responsible for the generation of the BAU. In the case of maximal $CP$
violation in Fig.~\ref{fig5}c, the phases $\alpha$ and $\beta$ are slightly
less constrained than in the previous case.

We have seen that in the present framework and assuming a mass hierarchy for
the Dirac neutrinos like the one of up-quarks (cf. Eq.~(\ref{gf14})), it is not
possible to reconcile the LMA, LOW and VO solar solutions with the leptogenesis
scenario. We may then ask ourselves what type of hierarchies should the Dirac
neutrino masses satisfy for these solutions to be realized and leptogenesis to
be viable. To answer this question let us now relax our previous assumption
(\ref{gf14}) and write the most general form for the Dirac neutrino mass
spectrum:
\begin{align} \label{num1}
d_D = m_{D3}\, \text{diag} (\varepsilon_1, \varepsilon_2, 1)\ ,
\end{align}
with $\varepsilon_i \equiv m_{Di}/m_{D3}$, $i=1,2$ and
$0 < \varepsilon_1 \leq \varepsilon_2 \leq 1$.

For a given scale of the heaviest Dirac neutrino $m_{D3}\ $, we can vary the
parameters $(\varepsilon_1,\varepsilon_2)$ in the above interval and look for
the allowed regions where the cosmological baryon asymmetry is within the
experimental range. To illustrate the results, we shall consider only the
presently most favoured solar solution, i.e. the LMA solar solution. The LOW
and VO solutions can of course be analyzed in a similar manner. The results are
plotted in Fig.~\ref{fig6} for two typical scales, namely, $m_{D3}$ of the
order of the bottom-quark mass $m_b \simeq 1$~GeV and around the top-quark mass
$m_t \simeq 100$~GeV at GUT scale. It is then clear from the figure that one
may obtain the correct baryon asymmetry, provided one uses a hierarchy for the
eigenvalues of $M_D$ corresponding to $\varepsilon_i$ lying inside the allowed
ranges indicated in Fig.~\ref{fig6}. However, note that for $M_D \propto M_u$
or $M_D \propto M_d$ one is not able to have the LMA solar neutrino solution
and yield the correct baryon asymmetry.

We emphasize that this analysis is quite general and does not rely on any
specific texture for Dirac matrix $M_D$ or the Majorana matrix $M_R$, except
for the fact that in the basis where the charged leptons and right-handed
Majorana neutrinos are diagonal, we assume a CKM-type misalignment between
$M_\ell$ and $M_D$.

\begin{figure*}
$$\includegraphics[width=8cm]{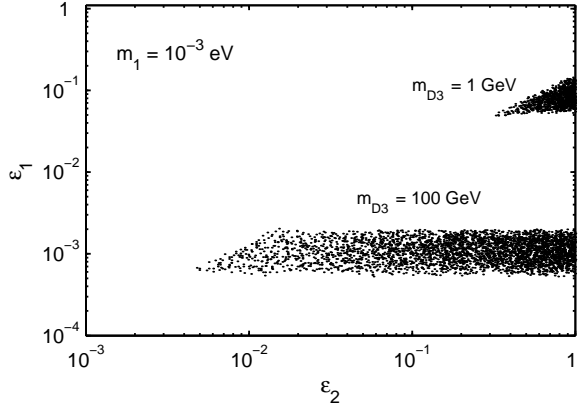}$$
\caption{Allowed region for the Dirac neutrino mass ratios $\varepsilon_i =
m_{Di}/m_{D3}$ in the case of the LMA solar solution. We assume $m_1 =
10^{-3}$~eV and two different scales for the heaviest Dirac neutrino, $m_{D3}
\sim m_b = 1$~GeV and $m_{D3} \sim m_t = 100$~GeV.} \label{fig6}
\end{figure*}

\section{Implications for low energy $CP$ violation and neutrinoless double beta decay }
\label{sec4}

It is clear from the analysis presented in the previous sections that $CP$
violation is a crucial ingredient for the generation of the cosmological baryon
asymmetry through leptogenesis. In our particular framework, the factor
$M_D^\dagger\,M_D$ is of the form
\begin{align} \label{cp1}
M_D^\dagger\,M_D = (U_R^\dagger\,d_D\,V_L)\,(V_L^\dagger\,d_D\,U_R) =
U_R^\dagger\,d_D^2\,U_R\ .
\end{align}
As a result, it does not depend directly on $V_L$, yet the presence of $V_L$ is
felt in the determination of $U_R$ through Eqs.~(\ref{gf16}) and also of the
heavy Majorana masses $M_i\ $ (i=1,2,3), which in turn appear in the lepton
asymmetry $\epsilon_{N_1}$. In Eq.~(\ref{cp1}) with $U_R$ written as
$P_\xi\,U'\,P_1$ (in the notation of Eq.~(\ref{gf11})) it is obvious that the
three phases in $P_\xi$ do not play any r\^{o}le for leptogenesis. Moreover, in
the limit $V_L = \openone$, we have
\begin{align} \label{cp2}
M_D = d_D\,U_R = P_\xi\,d_D\,U'\,P_1\ ,
\end{align}
and in this case $P_\xi$ can be rotated away from $M_D$ so that only three
physical $CP$-violating phases remain in $M_D$. In this case it is also
possible to choose a WB where both $M_D$ and $M_\ell$ are diagonal and real,
hence
\begin{align} \label{cp3}
M_R = U_R^\ast\,d_R\,U_R^\dagger\ ,
\end{align}
implying that all $CP$-violating effects could be generated at high
energies with $CP$ softly broken.

One of the striking features of our numerical results is the fact that in the
present framework leptogenesis favours a Dirac phase $\delta$ in the PMNS
mixing matrix $U_\nu$ very close to $\pi$ for relatively large values of
$|U_{e3}| \gtrsim 10^{-2}$. In the exact limit $\delta = \pi$ all
$CP$-violating effects at low energies are due to the presence of the Majorana
phases $\alpha$ and/or $\beta$. Nevertheless, the determination of the nature
of the phases relevant to the lepton asymmetry is not trivial and would require
a careful analysis of the matrix $R$ in Eq.~(\ref{gf8}). This is due to the
fact that $U_R$ and $U_\nu$ are related in a complicated manner even in the
limit $V_L = \openone$ as can be seen from Eqs.~(\ref{gf16}). Another
remarkable feature is the fact that the Majorana phases $\alpha$ and $\beta$
can only give rise to the necessary baryon asymmetry in those regions of the
parameter space represented by the bands in Figs.~\ref{fig5}, where one can see
that although all possible values of $\alpha$ and $\beta$ are allowed, they are
not independent of each other. The type of pattern generated in the
($\alpha,\beta$)-plane remains very similar for different values of the Dirac
phase $\delta$. It is also noticeable that leptogenesis is viable with a single
nonzero $CP$-violating phase of Majorana character even for $\delta = 0, \pi$,
whilst the Dirac phase alone is not sufficient.

The three phases $\delta, \alpha$ and $\beta$ contained in the mixing matrix
$U_\nu$ (for 3 generations of light neutrinos) are precisely the ones related
to $CP$-violating phenomena at low energies. It is well known that neutrino
oscillations are only sensitive to Dirac-type phases since the combination of
matrix elements of $U_\nu$ appearing in the oscillation probability is such
that the Majorana phases cancel out. $CP$ violation in oscillations will
manifest itself in the difference of $CP$-conjugated oscillation probabilities,
e.g.
\begin{align} \label{cp4}
P(\nu_e\rightarrow\nu_\mu)-P(\bar{\nu}_e\rightarrow\bar{\nu}_\mu)
\propto \mathcal{J}\ ,
\end{align}
with $\mathcal{J}$ a function of the PMNS matrix which can be expressed in
terms of experimentally relevant quantities as
\begin{align} \label{cp5}
\mathcal{J}=\frac{1}{4}\sin 2\,\theta_\odot \sin
2\,\theta_{\text{a}}\,(1-|U_{e3}|^{\,2})\,|U_{e3}|\,\sin\delta\,.
\end{align}

In the quark sector of the SM the parameter $\mathcal{J}$ defined in terms of
the $V_{CKM}$ elements is known to be $O(\lambda^6) \simeq 10^{-5}$. One might
expect that for large solar mixing $\mathcal{J}$ would be much larger, yet the
constraints imposed by leptogenesis in the context of SO(10) with the Dirac
neutrino masses fixed by the up quark spectrum imply an upper bound on
$\mathcal{J}$ of the order of $10^{-4}$, which is outside the experimental
reach of the next generation of neutrino experiments. We notice however that
within the LMA solution, which is at present the most favoured after the recent
SNO results, and allowing for the most general form for the Dirac neutrino
spectrum as in Eq.~(\ref{num1}), one can reach values of $\mathcal{J}$ of the
order of $10^{-2}$, thus rendering $CP$-violating effects visible in the near
future.

Neutrino oscillation experiments give no evidence about whether neutrinos are
Dirac or Majorana particles. Processes which violate lepton number such as
neutrinoless double beta decay ($(\beta\beta)_{0\nu}$-decay) of even-even
nuclei, $(A,Z)\rightarrow(A,Z+2)+e^-+e^-$, would imply that neutrinos have
Majorana character. The probability amplitudes of these processes are
proportional to the so-called ``effective Majorana mass parameter''
\begin{align} \label{ndbd1}
\meff =
\left|m_1\,U_{e1}^{\,2}+m_2\,U_{e2}^{\,2}+m_3\,U_{e3}^{\,2}\right|\ ,
\end{align}
(with the possibility of complex $U_{ei}$), or equivalently,
\begin{align} \label{ndbd2}
\meff = \left| m_1\,c_\odot^2\,(1-|U_{e3}|^2)+m_2\,s_\odot^2\,(1-|U_{e3}|^2)\,
e^{2i\alpha} + m_3\,|U_{e3}|^2\,e^{-2i\delta}\,e^{2i\beta}\right|\ .
\end{align}

Presently, the most stringent constraint on the parameter $\meff$ comes from
the $^{76}$Ge Heidelberg-Moscow experiment \cite{baudis}, indicating that
$\meff <\,$0.35 eV (90 \% C.L.). This value is still too high to provide some
information about the neutrino spectrum. In fact, it can only rule out some
models with quasi-degenerate Majorana neutrinos. Nevertheless, higher
sensitivity is planned to be reached in new generation experiments. The NEMO3
\cite{nemo} and CUORE \cite{cuore} experiments intend to achieve a sensitivity
up to $\meff \simeq 0.1$~eV while for the EXO \cite{exo} and GENIUS
\cite{genius} experiments this value is about one order of magnitude below.

A hierarchical spectrum with $m_1 \leq m_2 \simeq \sqrt{\dmsol} \ll m_3 \simeq
\sqrt{\dmatm}$ together with the CHOOZ bound on $|U_{e3}|$ would imply $\meff
\lesssim 10^{-3}$~eV, far below the present bound. On the other hand, for an
almost degenerate spectrum with $m = m_1 \simeq m_2 \simeq m_3\ $, for $m >
0.35$~eV some cancellation between the terms in (\ref{ndbd2}) is already
required, while in the case of small solar mixing, the contribution comes
mainly from the term proportional to $m_1\ $. Finally, for an
inverted-hierarchical spectrum where $m_3 \leq m_1 \simeq m_2 \simeq
\sqrt{\dmatm}\ $, the maximal possible value is $\meff \simeq \sqrt{\dmatm}
\sim 7 \times 10^{-2}$~eV \cite{bilenky}.

We conclude this section with the following remark. It has been recently
claimed \cite{klapdor} based on a refined analysis of the data of the
Heidelberg-Moscow experiment that there is evidence up to 3.1$\sigma$ for the
observation of neutrinoless double beta decay with $\meff = (0.05 - 0.84)$~eV
(at 95\% CL) with a best-fit value of 0.39~eV. If confirmed, this result would
imply an almost degenerate or inverted hierarchy for the light neutrino masses
in the case of three generations \cite{sarkar}. If this is the case,
baryogenesis via leptogenesis based on the simplest SO(10) GUT scenario would
not be a viable mechanism, since it is not possible to reconcile any of the
solar solutions with the required cosmological baryon asymmetry, as it follows
from our analysis.

\section{Conclusions}
\label{sec5}

We have studied leptogenesis in the framework of a simple extension of the SM
where the right-handed neutrinos are added to the standard spectrum, thus
leading to the seesaw mechanism. In order to restrict the number of free
parameters, we have made use of some GUT-inspired relations for the quark and
lepton mass matrices. We have shown that the latter relations, together with
the constraints from the low energy neutrino data, imply important restrictions
on the size of leptogenesis. In particular, we have pointed out that for the
Just-So$^2$ and SMA solar solutions, one can generate sufficient BAU through
leptogenesis even for the simplest SO(10) GUT. On the contrary, for the LMA,
LOW and VO solar solutions, a different hierarchy for the Dirac neutrino masses
is required in order to obtain a viable leptogenesis.

We expect our analysis and conclusions to remain valid also in the
supersymmetric version of the present framework. Although in this case new
decay channels will enhance the generated $CP$ asymmetry, these additional
contributions tend to be compensated by the washout processes which are in
general stronger than in the SM case \cite{plumacher}.

A related and very important subject is the search for $CP$ violation in the
leptonic sector, at low energies. This is at present one of the great
challenges in particle physics and it has recently received a great deal of
attention. Experiments with superbeams \cite{super} and neutrino beams from
muon storage rings (neutrino factories) \cite{factory} have the potential to
measure directly the Dirac phase $\delta$ through $CP$ and $T$ asymmetries
\cite{asym} or indirectly through oscillation probabilities which are
themselves $CP$ conserving but also depend on $\delta$ . An alternative method
proposed and discussed recently \cite{farzan} is to measure the area of
unitarity triangles defined for the leptonic sector \cite{aguilar}.

In this paper, we have investigated the possible connection between $CP$
violation at low energies and leptogenesis. In our SO(10) inspired framework,
once masses and mixing for three light neutrinos are fixed, the baryon to
entropy ratio $Y_B$ can be computed. It is well known that $CP$ violation at
low energies depends on three physical phases appearing in the PMNS matrix. One
of them, $\delta$ , is a Dirac-type phase to which neutrino oscillations are
sensitive, the other two are Majorana-type phases which are relevant to
processes such as neutrinoless double beta decay. We have shown that the
required cosmological baryon asymmetry can be produced with only one of the
Majorana-type phases different from zero, while the same is not possible with
only a nonvanishing Dirac-type phase. Concerning the important question of
whether the strength of $CP$ violation at low energies will be sufficient to be
measured through neutrino oscillations, we have verified that there are two
possible scenarios: if one assumes minimal SO(10) then, the constraints imposed
by leptogenesis render these effects too small to be seen in the next
generation of neutrino experiments; on the other hand, if one allows for a more
general spectrum of the Dirac neutrino mass matrix, then the strength of $CP$
violation can be sufficient to be visible at low energy neutrino oscillation
experiments.

The leptogenesis scenario crucially depends not only on the mechanism that was
responsible for populating the early Universe with right-handed neutrinos but
also on the precise details of the reheating process after inflation. In this
paper we have only considered the conventional scenario, namely, the decays of
right-handed Majorana neutrinos which are produced in thermal equilibrium
processes. Other production mechanisms may be as well viable and even
competitive \cite{giudice}. Their realization will of course depend on the
particular details of the inflationary epoch and the thermal evolution of our
Universe.

\vspace{1cm}

\textbf{Acknowledgements}

We are grateful to Martin Hirsch for useful comments. The work of R.G.F. and
F.R.J. was supported by {\em Funda\c{c}\~{a}o para a Ci\^{e}ncia e a Tecnologia}
(FCT, Portugal) under the grants SFRH/BPD/1549/2000 and PRAXISXXI/BD/18219/98,
respectively. G.C.B. and M.N.R. received partial support from FCT under the
project POCTI/36288/FIS/2000 and the projects CERN/P/FIS/40134/2000 and
CERN/FIS/43793/2001, and also from the European Commission under the RTN
contract HPRN-CT-2000-00149.

\end{document}